\documentclass[final,5p,twocolumn]{elsarticle}

\usepackage{amssymb}
\usepackage{amsmath}
\usepackage[caption=true]{subfig}
\usepackage{multirow}
\usepackage{algorithm}
\usepackage{enumerate}
\usepackage{algpseudocode}
\usepackage{float}
\usepackage[colorlinks=true,linkcolor=black, citecolor=blue, urlcolor=blue]{hyperref}
\usepackage{tikz,pgfplots}
\usepackage{breqn}
\usepackage{bbm}
\usepackage{cleveref}
\usepackage{mathtools}
\usepackage{balance}

\pgfplotsset{%
    compat=newest, 
    tick label style={font=\footnotesize},
    label style={font=\small},
    legend style={font=\small},
    axis x line = center,
    axis y line = center,
    every axis/.style={pin distance=1ex},
    trim axis left
    } 

\newcommand{\iu}{\mathrm{i}\mkern1mu}

\makeatletter
\def\ps@pprintTitle{%
  \let\@oddhead\@empty
  \let\@evenhead\@empty
  \def\@oddfoot{\reset@font\hfil\thepage\hfil}
  \let\@evenfoot\@oddfoot
}
\makeatother

\begin{document}

\begin{frontmatter}

\title{The Epps effect under alternative sampling schemes}

\author[uct-sta]{Patrick Chang}
\ead{patrick.chang@eng.ox.ac.uk}
\author[uct-sta]{Etienne Pienaar}
\ead{etienne.pienaar@uct.ac.za}
\author[uct-sta]{Tim Gebbie}
\ead{tim.gebbie@uct.ac.za}
\address[uct-sta]{Department of Statistical Sciences, University of Cape Town, Rondebosch 7700, South Africa}

\begin{abstract}
Time and the choice of measurement time scales is fundamental to how we choose to represent information and data in finance. This choice implies both the units and the aggregation scales for the resulting statistical measurables used to describe a financial system. It also defines how we measure the relationship between different traded instruments. As we move from high-frequency time scales, when individual trade and quote events occur, to the mesoscales when correlations emerge in ways that can conform to various latent models; it remains unclear what choice of time and sampling rates are appropriate to faithfully capture system dynamics and asset correlations for decision making. The Epps effect is the key phenomenology that couples the emergence of correlations to the choice of sampling time scales. Here we consider and compare the Epps effect under different sampling schemes in order to contrast three choices of time: calendar time, volume time and trade time. Using a toy model based on a Hawkes process, we are able to achieve simulation results that conform well with empirical dynamics. Concretely, we find that the Epps effect is present under all three definitions of time and that correlations emerge faster under trade time compared to calendar time, whereas correlations emerge linearly under volume time.
\end{abstract}


\end{frontmatter}

\section{Introduction}\label{sec:intro}

The Epps effect \cite{EPPS1979} is key phenomenology relating to high-frequency correlation dynamics in financial markets. Central to the effect is the relationship between correlations and the time scale of the sampling scheme. The Epps effect is prevalent and has been observed in a range of financial markets ranging from stock markets to foreign exchange markets (see \citet{MMZ2011} and references therein).

The Epps effect is a long-studied phenomenon that still lacks a satisfactory and comprehensive explanation. The main sources that contribute towards this effect include: (i) asynchrony, (ii) lead-lag, and (iii) tick-size (discretisation of financial prices). We highlight a few studies by various scholars over the last few decades. \citet{RENO2001} explored the effect of asynchrony under the presence of lead-lag. \citet{PI2007} demonstrated that different levels of asynchrony resulted in different behaviours of the Epps effect. \citet{TK2007} derived an analytical expression that characterised the Epps effect as a function of the rate parameter from Poisson sampling. They extended this further by decomposing the correlation at a certain time scale $\Delta t$ as a function of the correlation at smaller time scales $\Delta t_0$ \cite{TK2009}. The analytic expression was then further extended by \citet{MMZ2011} who separated the effect of asynchrony and that of lead-lag. Finally, \citet{MSG2010} investigated the direct impact of tick-size on the Epps effect. 

Another strand of literature around the Epps effect focuses on estimators correcting for this decay in correlation. The canonical estimator that addresses this is the Hayashi--Yoshida estimator \cite{HY2005} which corrected the effect of asynchrony by allowing for multiple contributions using a cumulative covariance estimator. \citet{BLPP2019} characterised the correction through the probability of flat trading whereas \citet{MSG2011} provided a correction that compensated for both the statistical cause of tick-size and asynchrony. Other notable estimators that address this include: the quasi-maximum likelihood estimator from \citet{AFX2010} and the multi-asset lagged adjustment estimator from \citet{BCP2020}.

Part of the reason why the phenomenon still lacks a satisfactory explanation is because these causes do not account for the entirety of the Epps effect \cite{MMZ2011,MSG2011,TK2009}. The Epps effect is commonly understood as a bias that requires a correction \cite{BLPP2019,ZHANG2010}. However, \citet{PCEPTG2020b} speculated whether the Epps effect is better understood as a fundamental property arising from the discrete nature of high-frequency financial markets, and questioned whether diffusion processes are an appropriate underlying model representation. Nonetheless, all the aforementioned studies investigated the Epps effect under calendar time sampling. \citet{PCRBTG2019} deviated from this by providing a preliminary investigation of the effect under volume time sampling.

Time and the choice of time scales is fundamental to how we choose to represent information and data. This implies both the units and the aggregation scales for the resulting statistical measurables used to describe financial systems \cite{CTPA2011a}. At low-frequency the measurement time scales conform well with those convenient to human operators and researchers intuition --- calendar time. However, as one moves into the world of heterogeneous inter-related individual events, before meaningful correlations have emerged, it becomes increasingly difficult to uniquely provide global measures of event synchronisation to the system. 

The first real attempt at linking the dynamics under different definitions of time was by \citet{DERMAN2002}. Derman was able to derive the relationship between different definitions of time by assuming that each stock has its own {\it intrinsic time} that has a linear map to calendar time based on the stock's trading frequency. Under these assumptions, Derman was able to show that correlations are the same in calendar and intrinsic time. This linear assumption is problematic because the inter-arrivals of empirical events (market orders) are not homogeneous, but rather follows a Weibull distribution \cite{CTPA2011a}.

The goal of this paper is to explore the Epps effect under alternative sampling schemes. This is similar to what has been done for the Realised Volatility (RV) \cite{Fukasawa2010,GO2008,Oomen2006}. We explore this by using the Malliavin--Mancino (MM) Fourier estimator \cite{MM2002,MM2009} to directly probe the relationship between measured correlations and time scale represented as frequencies: (i) the number of Fourier coefficients $N$ will define the sampling frequency, and (ii) the reconstruction frequency $M$. Here we are using integrated estimates and thus do not need to concern ourselves with the choice of $M$ \cite{PC2020}. We will contrast this with the well understood Hayashi--Yoshida (HY) estimator. This comparison is informative because both are designed to compute estimates in calendar time. However, the HY is based on the {\it a priori} assumption that the process has some underlying Brownian motion contribution and is constructed to be unbiased in this sense. The MM estimator has no such reliance on the specification of bias with respect to the underlying process.

Many estimation methods do not allow the precise extraction of the correlations as a direct function of the sampling frequency independent of assumptions relating to the underlying stochastic processes. Fourier methods provide the technology to achieve this. In addition, the Fourier method allows direct extraction of correlations at a particular sampling time scale. This will enable us to efficiently probe correlations at different high-frequency time scales.

There are important caveats, one of which is captured by volume time. Volume time is an example of when we cannot explicitly use sampling to map a definition of time aggregated from local events into some global calendar time that preserves event synchronisation in both representations. This is because the rate of trading of each underlying stock is not well-defined in calendar time which can lead to cases where trading assets may be synchronised in volume time, but only intermittently in calendar time.

To simulate the order book events, in particular transactions, we need a time labelled 3-tuple with time, price, and volume: $(t,p,v)$. Using the event times defined by the Hawkes process we are free to choose the distributional properties of the prices and volumes associated with events. We know empirically that we can relate price changes $\Delta p_t$ of transactions at some time $t$ to the volumes at that time $v_t$ using a price impact function: $\Delta p_t \propto f(v_t)$. We could exploit the price impact function to simulate a realistic order-book of transaction events \cite{CTPA2011a}. Here we argue that one does not need to link the volumes to the price changes through price impact to recover realistic Epps curves under the various sampling schemes considered. Perhaps surprisingly, the fine-to-coarse model introduced by \citet{BDHM2013a} conforms well with reality when combined with a power-law distribution for the volumes \cite{CTPA2011a}.

The paper is structured as follows. \Cref{sec:time} introduces the different temporal metrics. \Cref{sec:est} presents the estimators and their relationship to each other. \Cref{sec:exp} uses a multivariate Hawkes process to simulate an event-based price process which is used to investigate and compare calendar time sampling, transaction time sampling and a proxy for volume time sampling. \Cref{sec:emp} investigates and compares the various sampling methods for banking securities on the Johannesburg Stock Exchange using trade-and-quote data. Finally, \Cref{sec:conc} we conclude with some closing remarks.
The key contribution made is to demonstrate that the Epps effect emerges with different dynamics depending on the choice of time. In particular, volume time rescales the exponential emergence of correlation into a linear scale. 

\section{Temporal metrics}\label{sec:time}

In order to investigate the Epps effect under alternative sampling schemes, we need to evaluate the various definitions of time \cite{CTPA2011a,GO2008,Oomen2006,Toke2011}. First we have {\it calendar time} also known as {\it physical time}. This is the most natural way to perceive time as it is the way humans experience time at low-frequency and it is common to all investors and markets. It is also the most common approach used to compute statistical properties of financial time series because it can be defined without reference to the system generating the data. Even though the passage of time is most natural to human operators under this definition, it is not without challenges. These problems include: intraday seasonality, asynchrony across stocks, different and often overlapping market calendars and time zones. However, calendar time can allow coordinated sampling in a unique way independent of the data generation process. 

There are other methods to measure the passage of time, approaches that arise from the mechanics of the data generating processes, and the technologies used to manage and make decisions in financial systems. We can count events and use the event arrivals to increment a time. The most general being {\it event time}. For example, if we consider the full order book then one may increment event time by one unit when an order book event occurs. This may be a trade, a cancellation, an order modification or any order book event that changes the order book in some way. This is related to the idea of decision time, but here we will use {\it event time} and {\it decision time} interchangeably because practically any event in the order book is the result of a decision. Some prudence should be exercised, more generally in an Economic system this need not be the case, as many events can be truly random and exogenous relative to the strategic agents making decisions and taking actions that generate events. An outcome of using event time can be the ``smoothing'' of the data which is caused by bursts of activity being stretched out in the time series --- as we have one event per time unit \cite{CTPA2011a}. This can lead to the effect that under calendar time there is significant clustering, while under different choices of event time the clustering effect can be smoothed away.  

There can be different types of event times. For example, when considering transaction prices, one may use {\it trade time} or {\it transaction time}. Using this count, time is increased by one unit each time a unique transaction occurs. The advantage of this count is that limit orders do not affect the flow of time. Additionally, aggregational normality is more clear in trade time \cite{CTPA2011a}. However, this choice of time can exclude the many cascades and burst of order book changes that can in turn lead to the actual trading events. 

One can also use {\it tick time}. Using this count, time is increased by one unit each time the price changes. \citet{GO2008} found that for the realised volatility, tick time sampling is superior to transaction time sampling, which are both an improvement on calendar time sampling \cite{Oomen2006}, at least with respect to using a performance measure such as the mean squared error. There are many such possible choices for performance measures.

Finally, one may use {\it volume time} \cite{ELO2012B} to incorporate the volume of transactions. Using this count, time is increased by one unit for each unit of volume transacted each time a single share or instrument is exchanged. The advantage of this count is that stocks can naturally be synchronised, irrespective of their liquidity, directly in terms of sequence of comparable volumes traded. This is a practical data-science construct which is useful for the aggregation of real-time relative risk measures such as VPIN \cite{ELO2012B}.

However, mapping event times to calendar times can be problematic. For example, mapping volume time to a calendar or external clock does not always make sense because the mapping is highly nonlinear, intermittent and system dependent. In finance, an example where this can make sense could be the aggregation of transactions from trade time to bar data, so called OHLCV bars. These are typically aggregated at 1 minute, 5 minutes, and so on; it should be expected to require increments that may be correlated.\footnote{It should be noted that mappings from trade events to bar data are still nonlinear mappings, but ones that may well be possible to faithfully represent using approximations that combine stochastic time and fractional process \cite{GJR2018}. However, these are still defined using continuous time limit representations without explicit coarse-grained sampling, and so may not be natural to a system that is fundamentally discrete at high-frequency \cite{PCEPTG2020b}.} Moreover, trade time can also be problematic when one mixes data generating processes. For example, the aggregation of intraday transactions in trade time to low-frequency transactions sampled below one day. Here the data generating processes can be vastly incommensurate; intraday transactions represent prices arising from continuous-time double auctions and traders operating on these time scales. Whereas, market close prices are from an entirely different market structure, specifically closing auctions. These can be populated by different trading agents with different decision horizons and liquidity expectations. Sampling daily closing auction prices to even lower frequencies, such as weekly, monthly or even yearly, can also be problematic if key properties of the power spectrum are not preserved. 

Here we restrict ourselves to high-frequency and intraday day time scales, {\it i.e.,} to the high-frequency and mesoscale. Here the intermediate mesoscale is defined by time scales at which correlations are still emergent. For the purposes of our discussion, low-frequency finance has well-defined correlations that can be well represented by traditional latent models. Here we conceptually associate low-frequency with anything sampled from daily closing auction data where agent decisions are synchronised by the mechanism of the auction process.

Aside from what time scales can be appropriate for a given event process and the associated decisions being made from the resulting data, the question of whether these can be uniquely aggregated or mapped into various notions of global system time depends on how the correlations between these event processes emerge in the representations and the implicit filtering that takes place on the underlying data. Any choice of time scale implies bounds on what can be measured and represented by those measures being computed from the resulting filtered and sampled data. The choice of time carries implicit implications for the uncertainty relations that link various measurables, particularly under averaging and sampling --- the {\it cardinal theorems of interpolation}
\cite{Luke1999,Nyquist1928,Robertson1929,Shannon1949a,Shannon1998}. However, here it is the Whittaker-Shannon interpolation theorems and the link to the Gabor limit theorems that is probably of most interest. We may need to be particularly careful of the bandwidth versus time-limiting constraints \cite{Gabor1946} when making choices relating the sampling time scales near to the emergence of events themselves. The importance of this will not be discussed any further here, however, it is something that we would like to urge readers to remain cognisant of as it may provide constraints on the correlations and their dynamics as they emerge.

The fundamental question remains unanswered: What choice of time is appropriate when we have chosen a particular sampling time scale when computing cross-correlations between different assets with vastly different event generation processes? Correlations have yet to emerge at small time scales near individual events. Whereas, at larger time scales, the aggregation of events impacts the nature of cross-correlations and how one measures the cross-impacts and relationships between event generation processes.

\section{Estimators}\label{sec:est}

\subsection{Malliavin--Mancino}\label{subsec:MM}

\citet{MM2002,MM2009} introduced a Fourier estimator that expresses the Fourier coefficients of the volatility process using the Fourier coefficients of the price process $X^i_t = \ln(p^i_t)$, where $p^i_t$ is the generic asset price at time $t$ for asset $i$.

Following \citet{MM2009}, by rescaling the transactions times between $[0, T]$ to $[0, 2\pi]$ and using the \textit{Bohr} convolution product, we have that for all $k \in \mathbb{Z}$ and $N$ samples:
\begin{equation} \label{eq:Der:1}
    \begin{aligned}
      \mathcal{F}(\Sigma^{ij})(k) = \lim_{N \rightarrow \infty} \frac{2 \pi}{2N+1} \sum_{|s| \leq N} \mathcal{F}(dX^i)(s) \mathcal{F}(dX^j)(k-s),
    \end{aligned}
\end{equation}
where $\mathcal{F}(\ast)(\star)$ is the $\star$th Fourier coefficient of the $\ast$ process.

Using previous tick interpolation and a simple function approximation for the Fourier coefficients (see \cite{MM2009}), we obtain the Dirichlet representation of the integrated volatility/co-volatility estimator:\footnote{Note that $\iu \in \mathbb{C}$ is such that $\operatorname{Re}(\iu)=0$ and $\operatorname{Im}(\iu)=1$. It should not be confused with integer indices $i$, for example on the times $t^i_h$.}

\begin{equation} \label{eq:Der:2}
    \hat{\Sigma}^{ij}_{n,N} = \frac{1}{2N+1} \sum_{\substack{|s|\leq N \\ h=0,\ell=0}}^{n_i-1,n_j-1} e^{\iu s(t^j_{\ell} - t^i_h)} \delta_{i}(I_h) \delta_{j}(I_{\ell}),
\end{equation}
where $U^i = \{ t_h^i \}_{h=0}^{n_i}$ and $U^j = \{ t_{\ell}^j \}_{\ell=0}^{n_j}$ are the asynchronous transactions times observed between $[0,T]$ for each asset, and the price fluctuations are:
$$
\delta_{i}(I_h) = X^i_{t_{h+1}^i} - X^i_{t_{h}^i}, \quad \delta_{j}(I_{\ell}) = X^j_{t_{\ell+1}^j} - X^j_{t_{\ell}^j}
$$ 
for asset $i$ and $j$ respectively, and $n_i$ is the sample dimension for price $p^i$ and $n_j$ that of the price $p^j$, which \textit{a priori} can be different.

The advantage behind the estimator is its ability to investigate different time scales through the choice of $N$ \cite{PCEPTG2020a}. The conversion is given as:
\begin{equation} \label{eq:Der:3}
  N = \left\lfloor \frac{1}{2} \left( \frac{T}{\Delta t} - 1 \right) \right\rfloor.
\end{equation}
This means that there is no need to re-sample the raw observations onto a homogeneous grid using previous tick interpolation, such as in the case when using the RV estimator.

\subsection{Hayashi--Yoshida}\label{subsec:HY}

\citet{HY2005} introduced a cumulative covariance estimator defined as:
\begin{equation} \label{eq:Der:4}
\begin{aligned}
  \hat{\Sigma}^{ij}_{T}
  &= \sum_{h=1}^{\# U^i} \sum_{\ell=1}^{\# U^j} \left( X_{t_{h}^i}^i - X_{t_{h-1}^i}^i \right) \left( X_{t_{\ell}^j}^j - X_{t_{\ell-1}^j}^j \right) w_{h \ell},
\end{aligned}
\end{equation}
where
\begin{equation} \label{eq:Der:5}
    w_{h \ell}=\left\{\begin{array}{ll}
    1 & \text { if }\left(t_{h-1}^{i}, t_{h}^{i}\right] \cap\left(t_{\ell-1}^{j}, t_{\ell}^{j}\right] \neq \emptyset, \\
    0 & \text { otherwise. }
\end{array}\right.
\end{equation}
Here $U^i = \{ t_h^i \}_{h \in \mathbb{Z}}$ is the set of asynchronous transactions observed between $\left[0, T\right]$ for asset $i$ and $\# U^i = n_i$ denotes the cardinality of set $U^i$.

The advantage behind the estimator is its ability to ameliorate the statistical cause of the Epps effect arising from asynchrony. The weakness of this estimator is that it is unable to investigate different time scales, which is particularly problematic because asynchrony is only one of the various sources of the Epps effect \cite{PCEPTG2020b}.

\subsection{Realised Volatility}\label{subsec:relation}

Both the Malliavin--Mancino and Hayashi--Yoshida estimator were designed to compute estimates in calendar time, while dealing with the issue of asynchronous arrival of transactions. Both estimators overcome the need to re-sample the process onto a synchronous and homogeneous grid, for example, by using previous tick interpolation. However, given that we plan to re-sample the process under different definitions of time, it becomes useful to consider how the two estimators relate to the Realised Volatility (RV) estimator.

In the case of the Hayashi--Yoshida estimator, when $t_{h}^{i}$ and $t_{\ell}^{j}$ are synchronous and homogeneously spaced, we see that \cref{eq:Der:4} reduces to the RV estimator given as:
\begin{equation} \label{eq:Der:6}
\begin{aligned}
  \hat{\Sigma}^{ij}_{T}
  &= \sum_{h=1}^{n} \left( X_{t_{h}^i}^i - X_{t_{h-1}^i}^i \right) \left( X_{t_{h}^j}^j - X_{t_{h-1}^j}^j \right).
\end{aligned}
\end{equation}

In the case of the Malliavin--Mancino estimator, when $t_{h}^{i}$ and $t_{\ell}^{j}$ are synchronous, homogeneously spaced and $N = n/2$ (the Nyquist frequency in the synchronous case, where $n=n_i=n_j$), then it was found in \cite{PCRBTG2019} that the estimate were numerically the same as the RV estimate. This will also be demonstrated and discussed in \Cref{subsec:exp:VT}.

\section{Experiments}\label{sec:exp}

Here we are only interested in transactions, for this reason the remainder of the paper we will be using event time and transaction time interchangeably. 

We will be comparing two types of sampling for calendar and event time. First, when using the RV estimator, we will down sample a synchronous and homogeneous grid (obtained through previous tick interpolation) by skipping observations to investigate different time scales. For example, if we have 28,200 samples that correspond to one second intervals in calendar time. This means we would drop half the samples by sampling every second observation to achieve two second intervals in calendar time. Second, when using the Malliavin--Mancino estimator, we will be sampling different time scales through the Fourier domain by choice of $N$ for the particular time definition. 

The Malliavin--Mancino estimator presents several advantages. First, it does not need to drop observations to investigate different time scales. For example, suppose we have a total of 100 events. To investigate two unit time scale in event time, we would need to drop half the events when re-sampling for the RV estimator; whereas with the Malliavin--Mancino estimator, we would use the full set of samples, but compute fewer Fourier coefficients to achieve the two unit time scale using the relation in \cref{eq:Der:3}. 

Second, it has the ability to deal with asynchrony both in calendar and event time (see \Cref{subsec:exp:ET} demonstrating why event time can present asynchrony). Therefore, it does not require previous tick interpolation to synchronise the observations. One must however be mindful that as a consequence of the sampling theorem that the Malliavin--Mancino estimator can only investigate time scales larger than the smallest interval between two transactions (in both event and calendar time) to prevent aliasing.

Third, because it can deal with asynchrony, it uses the same fixed observations when investigating different time scales. This leads to more stable estimates at different time scales \cite{PC2020}. On the other hand, observations are dropped when sampling at different time scales for the RV estimator, therefore the time series changes slightly each time through the loss in resolution. These changes cause larger changes to the correlation estimate at each time scale, making the estimates less stable \cite{PC2020}.

With regards to the Hayashi--Yoshida estimator, we include it as a baseline for calendar and event time. This is because the estimator corrects for the statistical cause of the Epps effect arising from asynchrony. Therefore, it is of interest to see the baseline correlation level corrected for asynchrony under each definition of time.

In the case of volume time, sampling in the Fourier domain with the Malliavin--Mancino estimator is not possible. This is because the different time scales in volume time are determined by the number of samples obtained, which depends on the size of the volume bucket, which determines how many shares will be aggregated to obtain a sample. This aggregation changes the time series at different time scales --- {\it it is not just a loss in resolution from skipping observations}.

\subsection{Hawkes process}\label{subsec:exp:hawkes}

To generate a price process based on events, we will be using the class of multivariate point processes introduced by \citet{HAWKES1971} known as a {\it Hawkes process}. Concretely, we will use the fine-to-coarse model introduced by \citet{BDHM2013a}. Let the bivariate log-price be:
\begin{equation}\label{eq:7}
\begin{aligned}
    &X_t^1 = X_0^1 + N_1(t) - N_2(t), \\
    &X_t^2 = X_0^2 + N_3(t) - N_4(t),
\end{aligned}
\end{equation}
where $\{N_m(t)\}_{m=1}^M$ is a 4-dimensional ($M=4$) mutually exciting Hawkes process with the associated intensity $\lambda(t) = \{\lambda^m(t)\}_{m=1}^M$ taking the form:
\begin{equation}\label{eq:8}
    \lambda^m(t) = \mu + \sum_{n=1}^M \int_{-\infty}^t \phi^{mn}(t-s) dN_s^n.
\end{equation}
The counting processes are coupled through:
\begin{equation}\label{eq:9}
    \boldsymbol{\Phi} = 
\begin{pmatrix}
0 & \phi^{(r)} & \phi^{(c)} & 0   \\
\phi^{(r)} & 0 & 0 & \phi^{(c)}   \\
\phi^{(c)} & 0 & 0 & \phi^{(r)}   \\
0 & \phi^{(c)} & \phi^{(r)} & 0
\end{pmatrix},
\end{equation}
where $\phi^{(r)} = \alpha^{(r)} e^{-\beta t}\mathbbm{1}_{t \in \mathbb{R}^+}$ and $\phi^{(c)} = \alpha^{(c)} e^{-\beta t}\mathbbm{1}_{t \in \mathbb{R}^+}$. The parameters are $(\mu, \alpha^{(r)}, \alpha^{(c)}, \beta) = (0.015, 0.023, 0.05, 0.11)$, borrowed from \citet{BDHM2013a}.

The interpretation of \cref{eq:9} is as follows: $\phi^{(r)}$ achieves mean reversion, because an uptick in $X^1$ by $N_1$ will lead to an increased intensity in the down tick $N_2$ --- allowing the price level to revert (similarly for $X^2$ through $N_3$ and $N_4$). While $\phi^{(c)}$ induces a correlation between the prices by connecting the two prices, since an uptick in $X^1$ by $N_1$ will lead to an increased intensity in the uptick of $X^2$ through $N_3$ (similarly for down ticks through $N_2$ and $N_4$).\footnote{The specification of \cref{eq:9} in the model is fully symmetric and thus there are no lead-lag effects (see \cite{BDHM2013a} and Remark 9 of \cite{BDHM2013b}).}

To ensure stability with stationary increments, the spectral radius of the branching matrix $\boldsymbol{\Gamma}$ must be strictly less than 1, {\it i.e.,} the magnitude of the largest eigenvalue must be less than 1. For our exponential kernel, we have $\boldsymbol{\Gamma} = \{ {\alpha^{mn}}/{\beta^{mn}} \}_{m,n=1}^M$.

A key insight from \citet{BDHM2013a} is their formulation of the covariance matrix for the model under calendar time sampling with increment size $\Delta t$:
$$
\begin{aligned}
\frac{C_{\Delta t}^{11}}{\Delta t}=& \Lambda+\frac{R C_{1}}{2 G_{1}}+\frac{R C_{2}}{2 G_{2}} \\
&+R \frac{C_{2} G_{1}^{2} \mathrm{e}^{-\Delta t G_{2}}-C_{1} G_{2}^{2}+Q_{1} G_{2}^{2} \mathrm{e}^{-\Delta t G_{1}}-C_{2} G_{1}^{2}}{2 G_{2}^{2} G_{1}^{2} \Delta t},
\end{aligned}
$$
and
$$
\begin{aligned}
\frac{C_{\Delta t}^{12}}{\Delta t}=& \frac{-R C_{1}}{2 G_{1}}+\frac{R C_{2}}{2 G_{2}} \\
&+\frac{R\left(C_{1} G_{2}^{2}-C_{2} G_{1}^{2}-C_{1} G_{2}^{2} \mathrm{e}^{-G_{1} \Delta t}+C_{2} G_{1}^{2} \mathrm{e}^{-G_{2} \Delta t}\right)}{2 G_{2}^{2} G_{1}^{2} \Delta t}.
\end{aligned}
$$
Here the parameters are:
$$
\begin{aligned}
\Lambda &= \frac{\mu}{1 - \Gamma_{12} - \Gamma_{13}},   \\
R &=\frac{\beta \mu}{\Gamma_{12}+\Gamma_{13}-1},    \\
C_{1} &=\frac{\left(2+\Gamma_{12}+\Gamma_{13}\right)\left(\Gamma_{12}+\Gamma_{13}\right)}{1+\Gamma_{12}+\Gamma_{13}},  \\
C_{2} &=\frac{\left(2+\Gamma_{12}-\Gamma_{13}\right)\left(\Gamma_{12}-\Gamma_{13}\right)}{1+\Gamma_{12}-\Gamma_{13}},  \\
Q_{1}=Q_{4}&=\frac{-\mu\left(\Gamma_{12}^{2}+\Gamma_{12}-\Gamma_{13}^{2}\right)}{\left(\left(\Gamma_{12}+1\right)^{2}-\Gamma_{13}^{2}\right)\left(1-\Gamma_{12}-\Gamma_{13}\right)},   \\
Q_{2}=Q_{3}&=\frac{-\mu \Gamma_{13}}{\left(\left(\Gamma_{12}+1\right)^{2}-\Gamma_{13}^{2}\right)\left(1-\Gamma_{12}-\Gamma_{13}\right)},
\end{aligned}
$$
and 
$$
G_{1}=\beta\left(1+\Gamma_{12}+\Gamma_{13}\right), \quad G_{2}=\beta\left(1+\Gamma_{12}-\Gamma_{13}\right).
$$
The correlation is then given by:
\begin{equation}\label{eq:10}
    \rho_{\Delta t}^{12} = \frac{C_{\Delta t}^{12}}{C_{\Delta t}^{11}},
\end{equation}
and the Epps effect is present.
Moreover, in the limit we have that \citep{BDHM2013a}: 
\begin{equation}\label{eq:11}
    \lim_{\Delta t \to \infty} \rho_{\Delta t}^{12} = \frac{2 \Gamma_{13}\left(1+\Gamma_{12}\right)}{1+\Gamma_{13}^{2}+2 \Gamma_{12}+\Gamma_{12}^{2}}.
\end{equation}

The experiments that will follow in the next few subsections are conducted by simulating 100 pairs of \cref{eq:7} for $T = 72,000$ seconds. Each of the price pairs will be re-sampled into the appropriate format and correlations will be computed under each definition of time for sampling intervals between 1 to 100 units. The mean correlation estimate at each sampling interval will be plotted along with error ribbons that contain 95\% of the estimates at each sampling interval. These are computed from the 100 replications using the student $t$-distribution with 99 degrees of freedom and the standard deviation of the estimates between the replications at each sampling interval.

Moreover, to demonstrate how the sampling is achieved under each definition of time, we simulate a single pair of \cref{eq:7} for $T=300$ seconds.

\subsection{Calendar time}\label{subsec:exp:CT}

\begin{figure*}[htb]
    \centering
    \subfloat[Sampled for RV]{\label{fig:CalendarTimePrices:a}\includegraphics[width=0.5\textwidth]{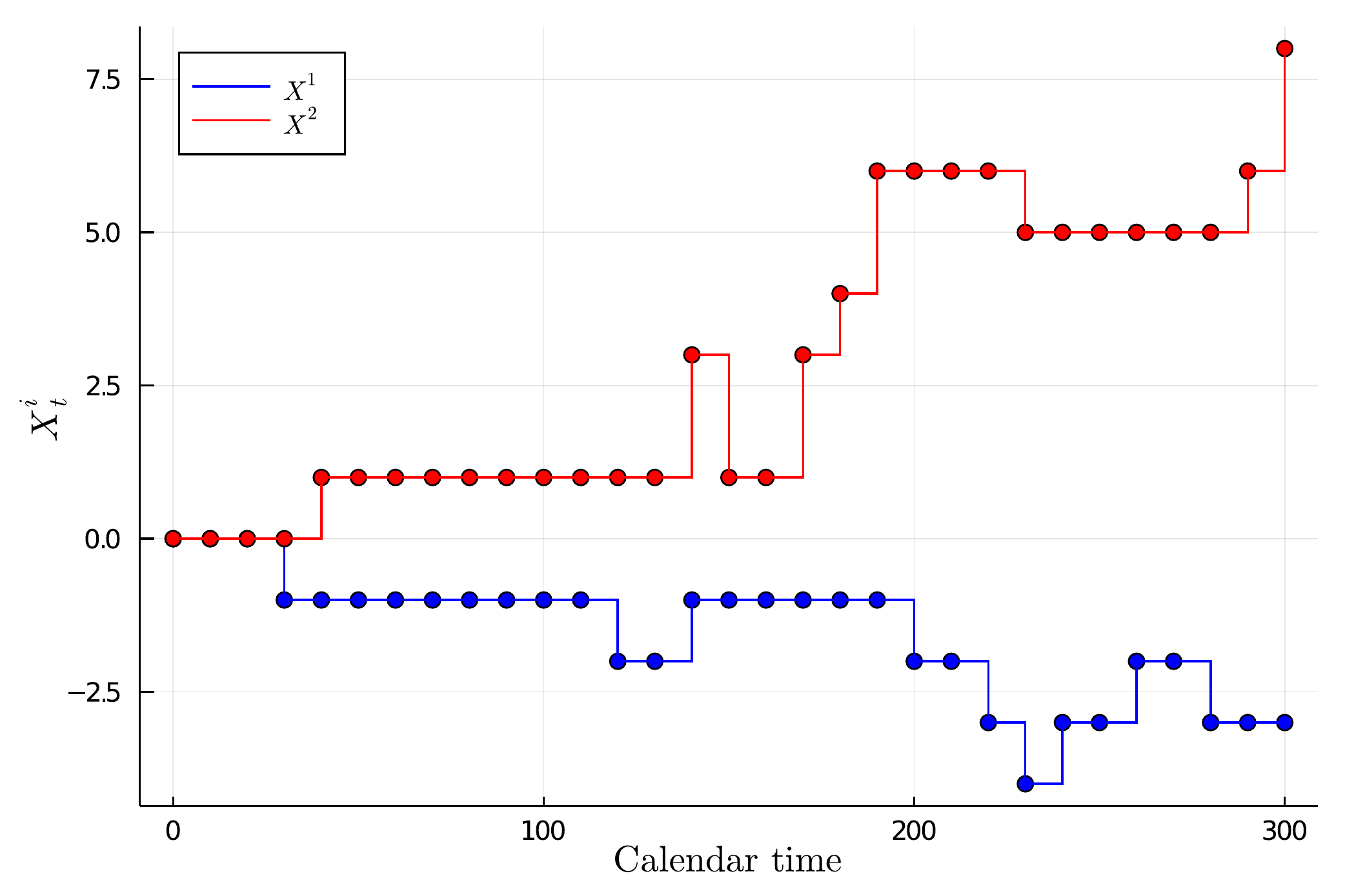}}
    \subfloat[Sampled for MM and HY]{\label{fig:CalendarTimePrices:b}\includegraphics[width=0.5\textwidth]{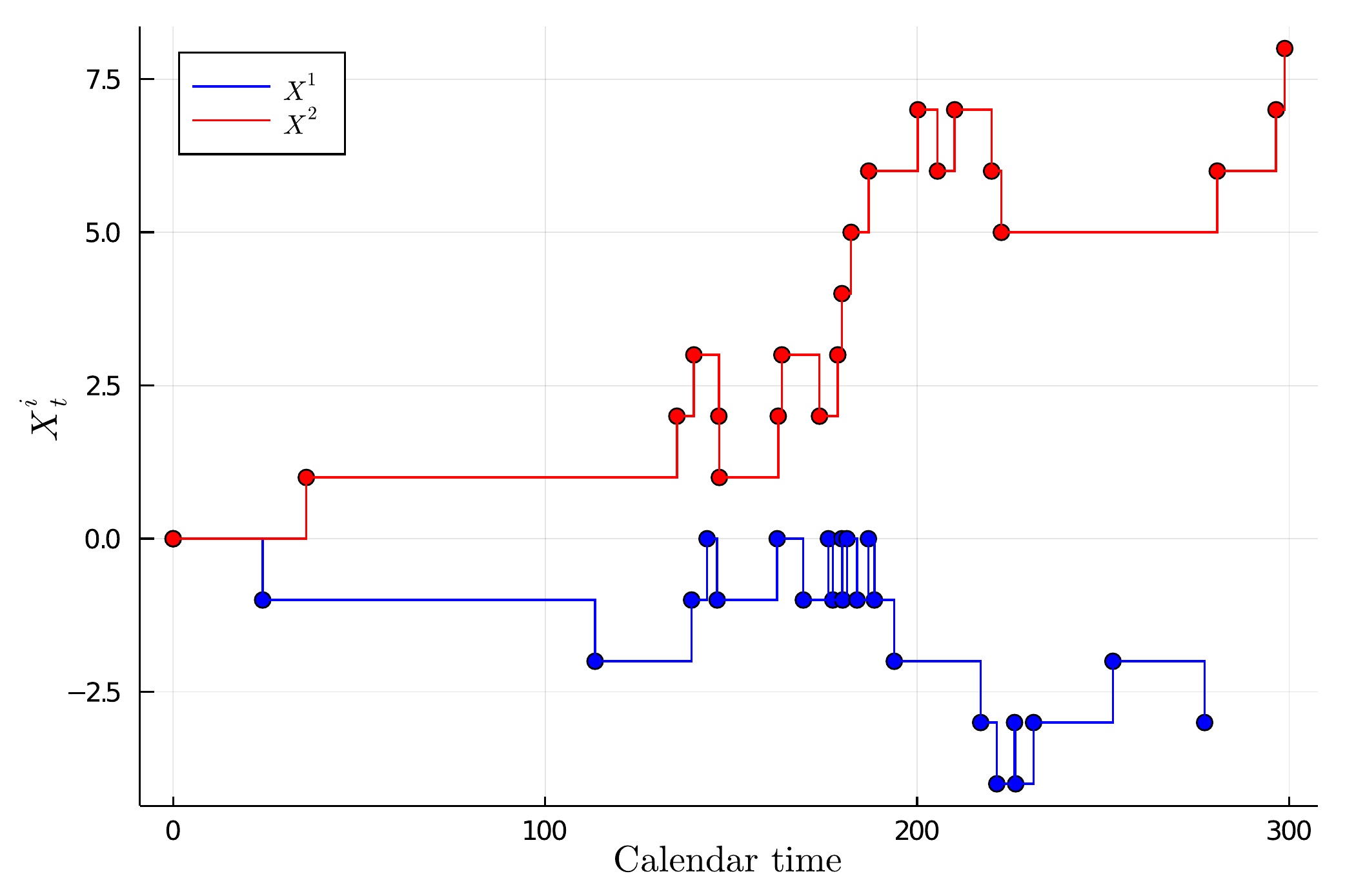}}
    \caption{The Hawkes price model in calendar time. The figure demonstrates the two types of input samples (given as bubbles) for the different estimators. (a) has synchronous and homogeneous samples with 10 unit intervals between observations. The synchronisation is achieved using previous tick interpolation. (b) has asynchronous samples based on when the events occurred.}
\label{fig:CalendarTimePrices}
\end{figure*}

\begin{figure}[!h]
    \centering
    \includegraphics[width = 0.5\textwidth]{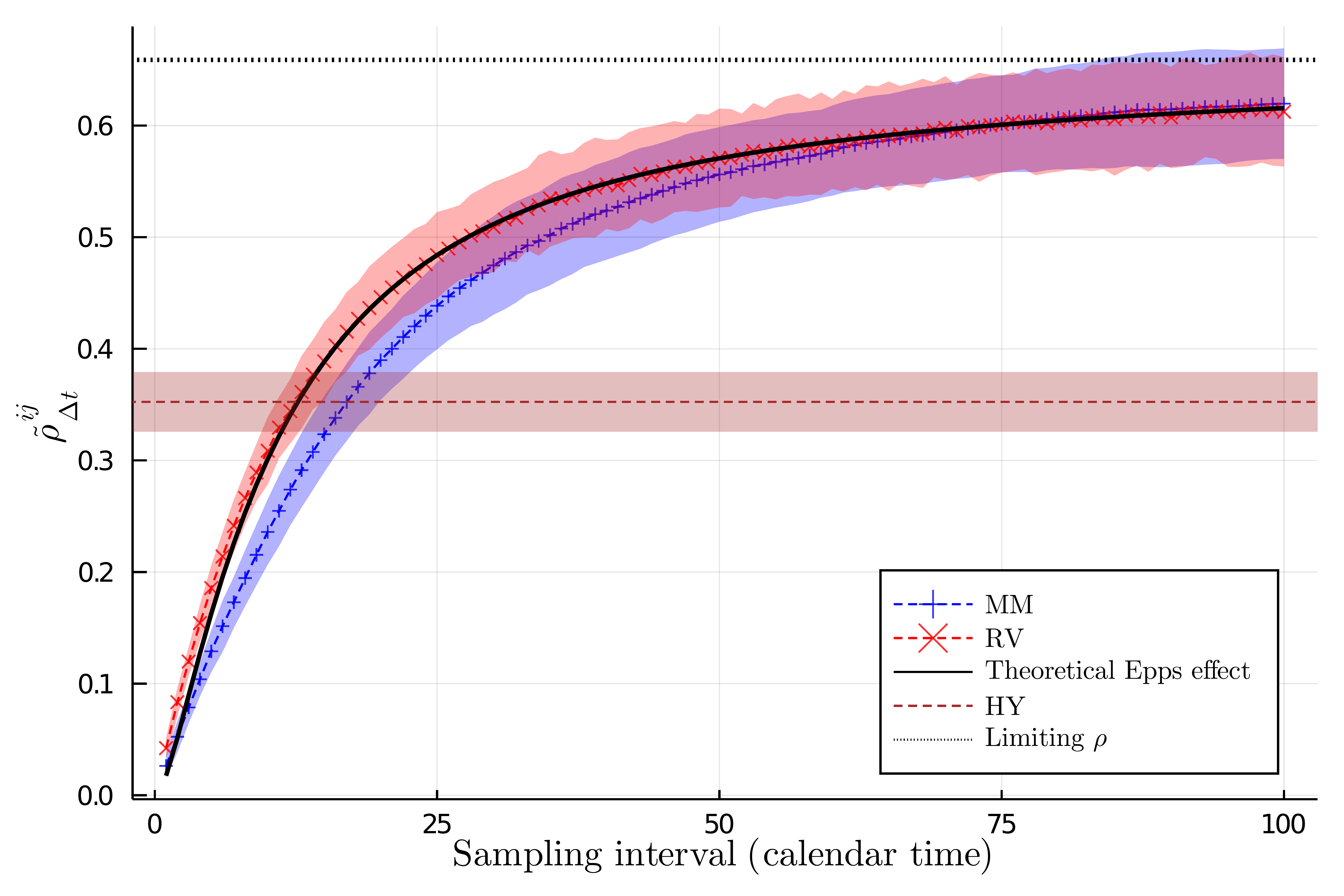}
    \caption{The correlation estimate at different sampling intervals in calendar time for the MM (blue) and RV (red) estimates. The HY (brown) estimate is provided as a baseline. The theoretical Epps effect from \cref{eq:10} and the limiting correlation from \cref{eq:11} are also provided.}
    \label{fig:SimCT}
\end{figure}

The calendar time sampling is relatively straightforward. When sampling for the RV estimator, we sample \cref{eq:7} at equidistant and synchronous points between $[0,T]$ using a particular sampling interval. As an example, \Cref{fig:CalendarTimePrices:a} demonstrates the samples (given as bubbles) where the sampling the done using 10 unit intervals (corresponding to 10 seconds). For the Malliavin--Mancino (MM) and Hayashi--Yoshida (HY) estimator, the samples are when the events (in this case jumps) occurred. This is demonstrated in \Cref{fig:CalendarTimePrices:b} where the bubbles are the samples. Notice that when the sampling interval increases for \Cref{fig:CalendarTimePrices:a} we lose more resolution into the price path which is present in \Cref{fig:CalendarTimePrices:b}.

\Cref{fig:SimCT} plots the correlation for the RV and MM estimates in calendar time measured at different sampling intervals with the HY estimates provided as a baseline estimate. 

We see that the RV estimator recovers the theoretical Epps effect which is expected as this was the estimator used to derive the covariance matrix in \citet{BDHM2013a}. The MM estimator does not quite recover the same theoretical curve but it captures a more pronounced the Epps effect. This is because there is the effect of asynchrony contributing towards the Epps effect in this case. The HY estimates recover a rather strange estimate. This is because it was found that the correction achieved here depends on the inter-arrivals of the asynchronous events \cite{PCEPTG2020b}. This fact was used by \citet{PCEPTG2020b} to design experiments to detect if the system has emerging correlations as in the case of \cref{eq:7} or if the system has correlations that exist at all sampling intervals such as a geometric Brownian motion.

\subsection{Event time}\label{subsec:exp:ET}

\begin{figure*}[htb]
    \centering
    \subfloat[Sampled for RV]{\label{fig:EventTimePrices:a}\includegraphics[width=0.5\textwidth]{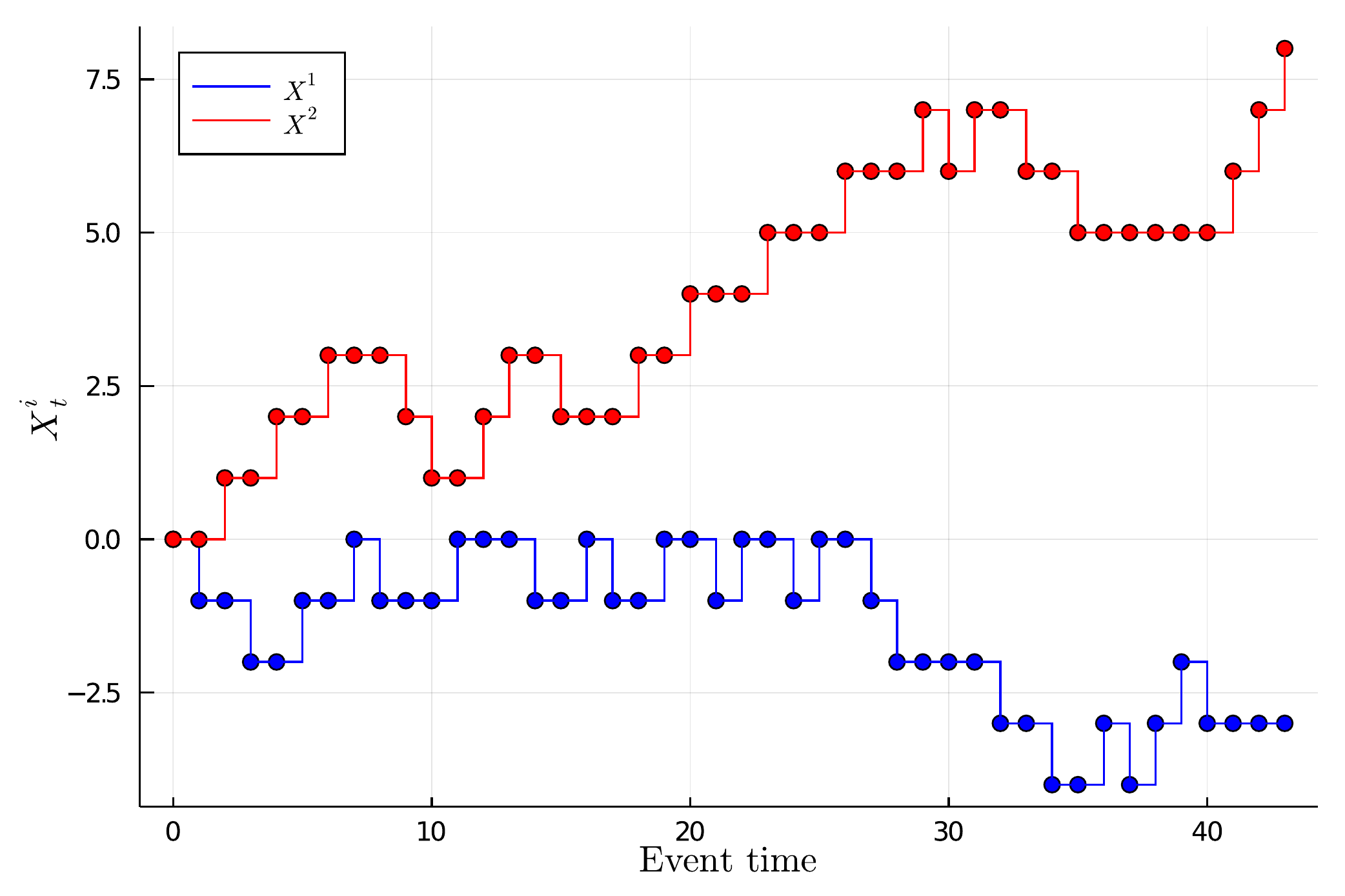}}
    \subfloat[Sampled for MM and HY]{\label{fig:EventTimePrices:b}\includegraphics[width=0.5\textwidth]{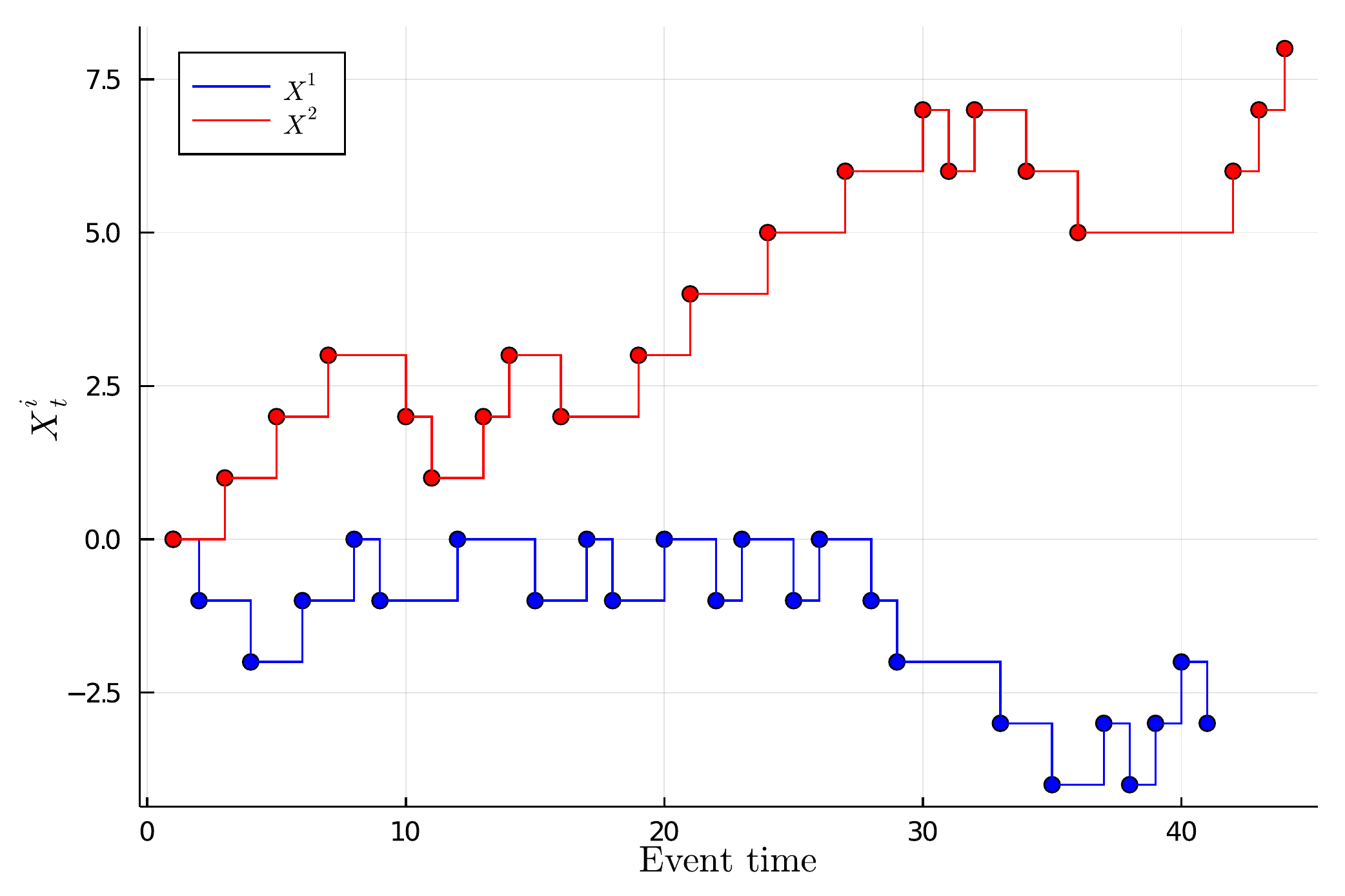}}
    \caption{The Hawkes price model in event time where the securities share the same event clock. The figure demonstrates the two types of input samples (given as bubbles) for the different estimators. (a) has synchronous and homogeneous samples with one unit intervals between observations. The synchronisation is achieved using previous tick interpolation. (b) has asynchronous samples on a homogeneous grid based on the ordering of when the events occurred.}
\label{fig:EventTimePrices}
\end{figure*}

\begin{figure}[!h]
    \centering
    \includegraphics[width = 0.5\textwidth]{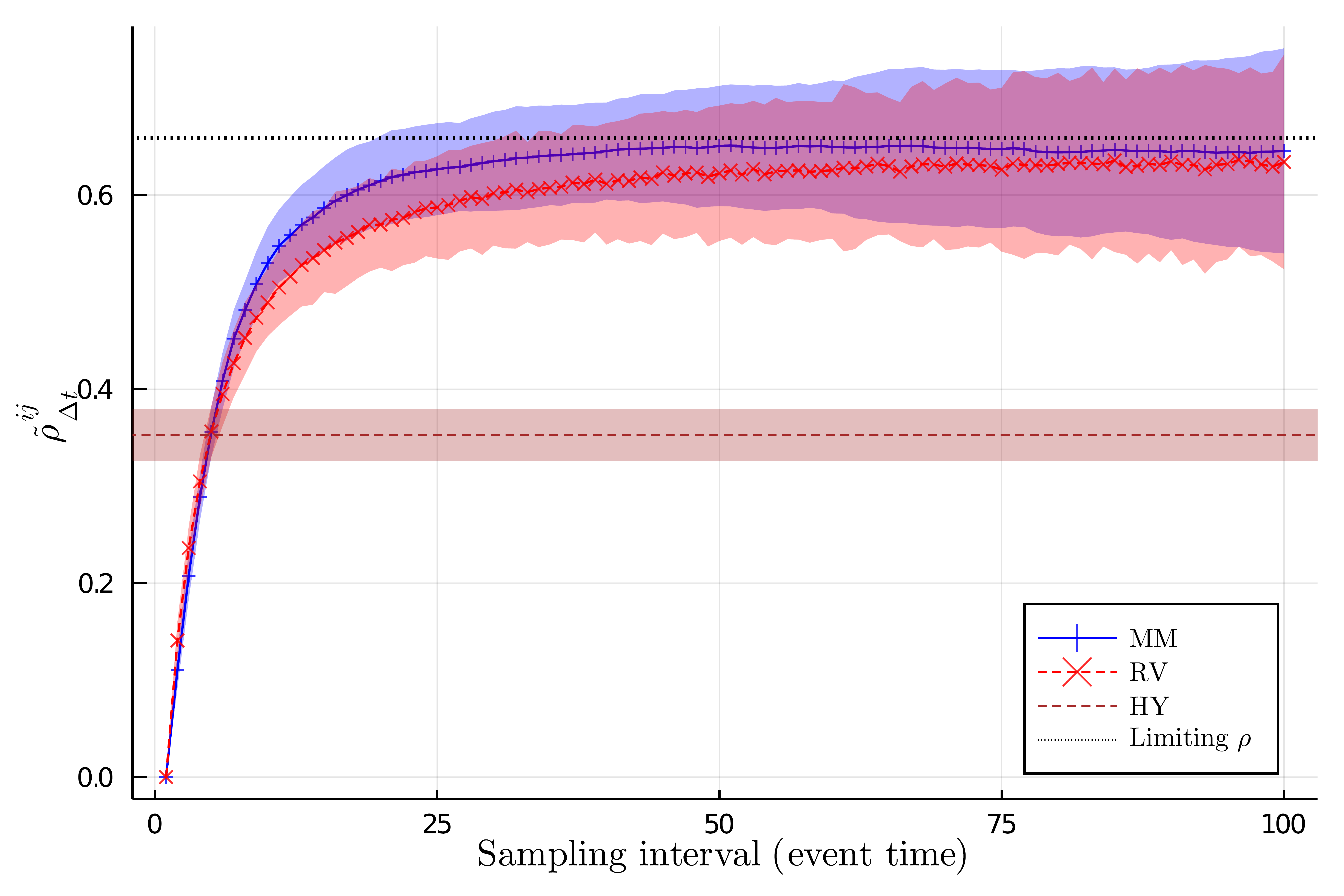}
    \caption{The correlation estimate at different sampling intervals in event time for the MM (blue) and RV (red) estimates. The HY (brown) estimate is provided as a baseline.}
    \label{fig:SimET}
\end{figure}

Event time is interesting because in theory, there is a one-to-one mapping between the event count $h$ to calendar time $t_h$ when considering one security.\footnote{This mapping can be difficult to find because there is often measurement error associated with when events occur on an exchange.} Event time however becomes slightly problematic when we want to compute the correlations between two securities. This is because each security will have a different number of events, therefore the event clock between securities are not comparable. Thus we need a method to unify the time between the securities. This can be achieved if the securities share the same event clock, where an event in either security will increase the time by one unit. 

Under this construction, the time series will be asynchronous but homogeneous, therefore we need to use previous tick interpolation to create the time series in event time for the RV estimator. \Cref{fig:EventTimePrices:a} demonstrates the samples in event time for the RV estimator where the sampling is done using one unit intervals. \Cref{fig:EventTimePrices:b} demonstrates the samples in event time for the MM and HY estimator where the samples are asynchronous but live on a homogeneous grid. Comparing \Cref{fig:EventTimePrices} against \Cref{fig:CalendarTimePrices:b}, we see that event time indeed achieves a smoothing effect as the bursts of activity in calendar time is stretched out.

\Cref{fig:SimET} plots the correlation for the RV and MM estimates in event time measured at different sampling intervals with the HY estimates provided as a baseline estimate.

We see that both the MM and RV estimates present an Epps effect in event time. Moreover, we see that correlations seem to emerge faster with a more pronounced concavity in the curves. The HY estimates in this case is particularly interesting, it actually recovers the same estimates as in calendar time as a result of how the event time is constructed (see a clearer comparison in \Cref{fig:SimComp}). This is because even though the time series is stretched out, the observations used in \cref{eq:Der:4} remain the same and because the events are asynchronous, the intervals where $w_{h \ell}$ is activated remains the same.

The relationship between the MM and RV estimates under the two definitions of time remains unclear. Particularly for the MM estimator. This is because the shifting of events through stretching out the time series will alter the power spectrum. Thus the relationship is unclear. Related to this, the pronounced concavity in event time could potentially be related to the shifting and stretching of the time series which could induce potential lead-lag effects. However, currently it is unclear how this mechanism could potentially explain the observed dynamics. The difficulty lies in that lead-lag is usually induced from a calendar time perspective through the shifting of observations (see the model in \citet{MMZ2011}), whereas here the time series is also stretched. Moreover, the shifting and stretching here is contingent on the asynchronicity between the two time series. Nonetheless, the well understood concave Epps effect remains present under both definitions of time.

\subsection{Volume time}\label{subsec:exp:VT}

\begin{figure*}[htb]
    \centering
    \subfloat[10 unit sampling interval in volume time]{\label{fig:VolumeTimePrices:a}\includegraphics[width=0.5\textwidth]{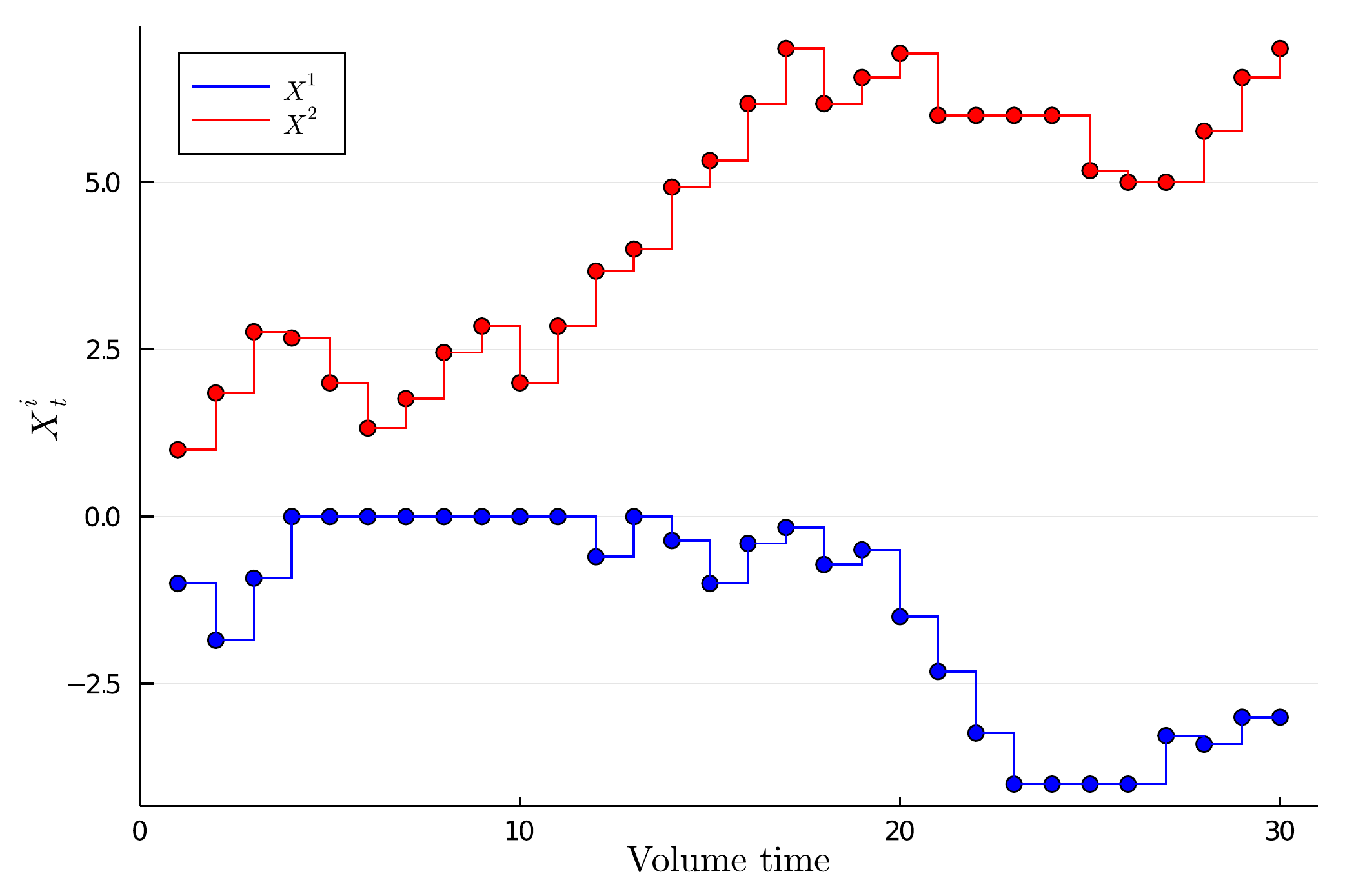}}
    \subfloat[5 unit sampling interval in volume time]{\label{fig:VolumeTimePrices:b}\includegraphics[width=0.5\textwidth]{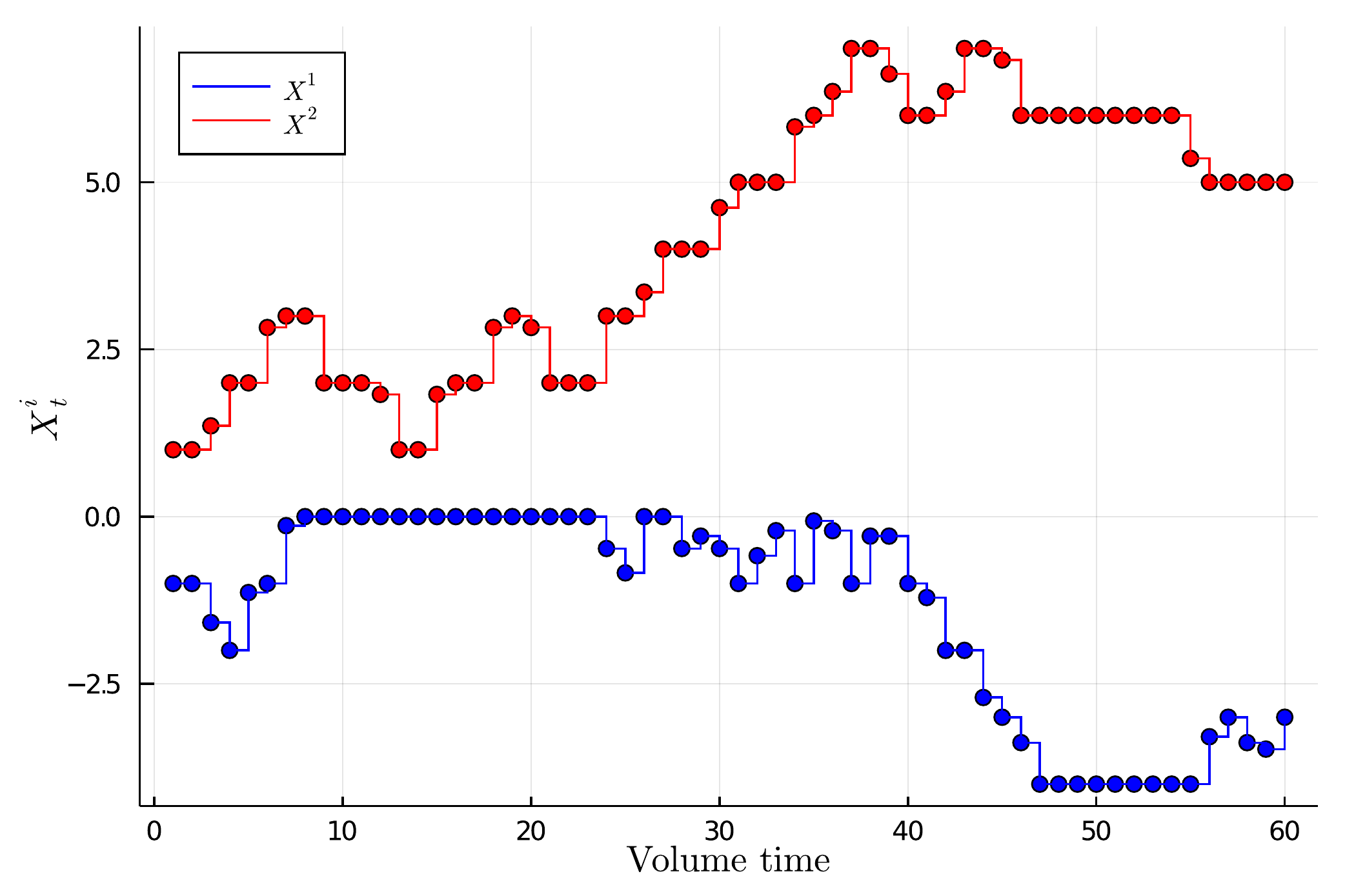}}
    \caption{The Hawkes price model with volumes samples from an IID power-law distribution given in \cref{eq:14}. The sampling interval was chosen to be the equivalent of (a) 10 unit and (b) 5 unit sampling interval in calendar time based on the number of samples obtained (relative to \Cref{fig:CalendarTimePrices}). The samples in volume time are synchronous, homogeneous and requires different sized volume buckets (for averaging) to achieve.}
\label{fig:VolumeTimePrices}
\end{figure*}

\begin{figure}[!h]
    \centering
    \includegraphics[width = 0.5\textwidth]{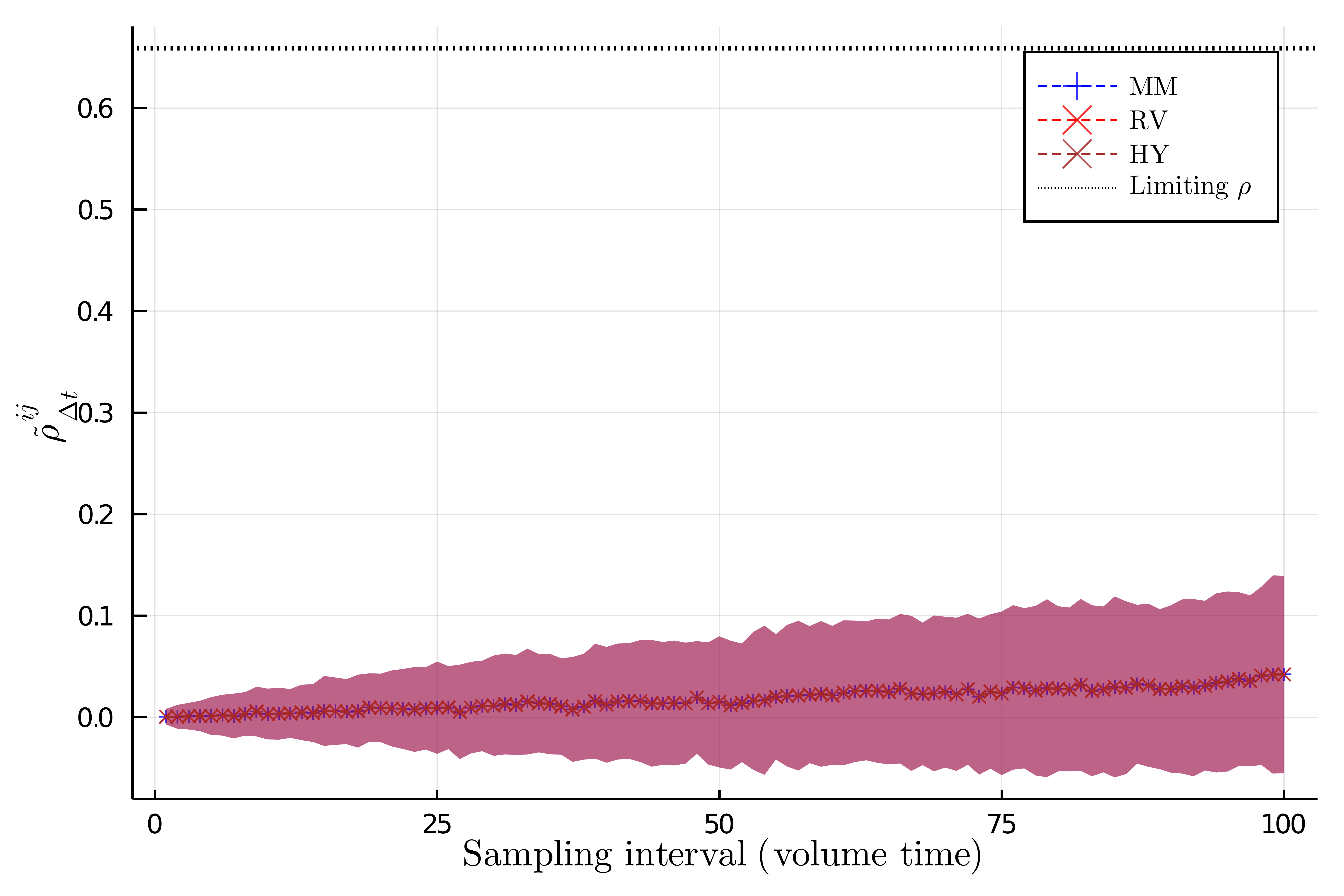}
    \caption{The correlation estimate at different sampling intervals in volume time for the MM (blue), RV (red) and HY (brown) estimates.}
    \label{fig:SimVT}
\end{figure}

\begin{figure}[!h]
    \centering
    \includegraphics[width = 0.5\textwidth]{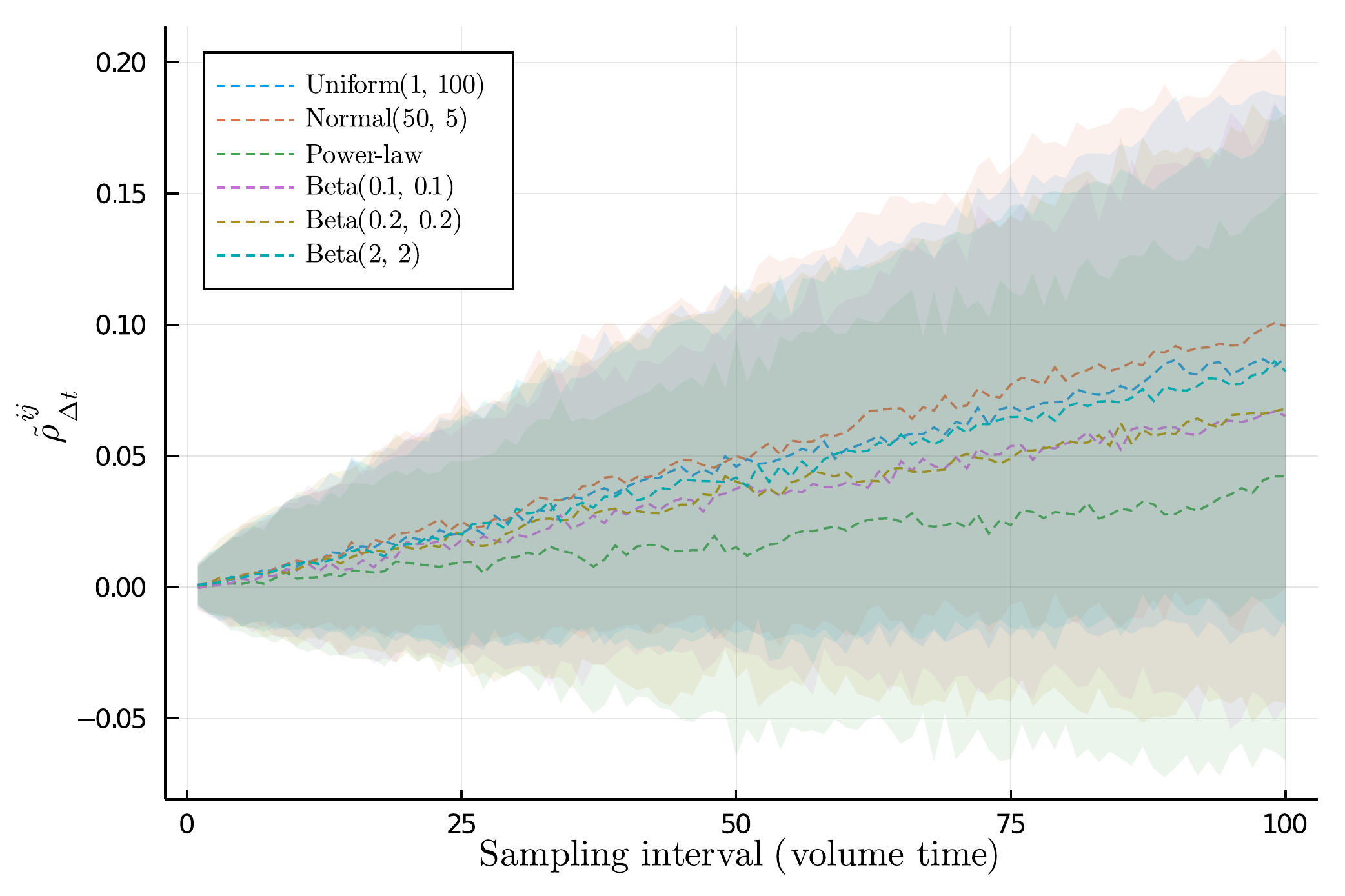}
    \caption{The Epps effect in volume time for different distributions generating the volume samples. The estimates are obtained using the RV estimator.}
    \label{fig:MoreVT}
\end{figure}

The sampling of volume time is different to that of calendar and event time. Although time in this definition is increased by one unit for each share traded, sampling it requires the use of volume buckets which determine the number of samples we get for each security. This method of sampling follows \citet{ELO2012B} which they used to compute VPIN. The sampling overcomes the problem of different volumes traded between securities. The problem however with this sampling method and volume time in general is that there is no clear mapping back to calendar time. This is because prices between transactions will be aggregated and multiple volume times will map back to the same calendar time. For example, suppose a transaction occurred with five shares traded. These five volume time counts would map back to the same instant in calendar time.

To obtain a notion of sampling interval in volume time, we will use the number of samples obtained to get an idea of the equivalent sampling interval in calendar time. For example, if a sampling interval of 10 units translates to 30 samples in calendar time, in order to obtain a 10 unit sampling interval in volume time, we need to pick a volume bucket that allows us to obtain 30 samples.

The sampling in volume time follows partly from appendix A1 from \citet{ELO2012B}. The transaction time series for a security $i$ is defined by the tuple:
\begin{equation}\label{eq:12}
(t^i_h, p^i_{t^i_h}, v^i_{t^i_h}),
\end{equation}
where $t^i_h$ is the transaction time, $p^i_{t^i_h}$ is the transaction price and $v^i_{t^i_h}$ is the transaction volume. Let $n$ be the number of samples we want from the volume time sampling, then the volume bucket $V$ is given as:
\begin{equation}\label{eq:13}
    V = \left\lfloor V_{\text{tot}} / n \right\rfloor,
\end{equation}
where $V_{\text{tot}} = \sum_h v^i_{t^i_h}$ is the total number of shares traded. Expand the number of observations by repeating each transaction $p^i_{t^i_h}$ as many times as $v^i_{t^i_h}$, generating $V_{\text{tot}}$ number of observations. Looping through the expanded prices, create a sample in volume time after $V$ shares by averaging the expanded prices, resulting in a total of $n$ samples. In other words, average $p^i_{t^i_h}$ by its volume contribution to each bucket $V$ for each volume time sample. Note that the last bucket is always incomplete or empty, thus these observations are discarded \cite{ELO2012B}. Under this sampling, we will have synchronous and homogeneous samples as each security will always have $n$ samples.

Crucially, Hawkes processes only model when the events occur; it does not provide us with information about the states of the features associated with these event times. Additional model assumptions are necessary to generate a realistic tuple.

A realistic model for the time series of transaction requires the model to account for the couplings between the tuples. First, the prices $p^i_{t^i_h}$ and volumes $v^i_{t^i_h}$ are coupled through market impact \cite{BGPW2004,LFM2003}.\footnote{Strictly speaking, market impact is the coupling between the mid-price $m_t$ and volume. However, the transaction price $p_t$ is related to the mid-price as: $p_t = m_t + s_t/2$ for buyer-initiated transactions where $s_t$ is the spread. Therefore, prices and volumes are coupled in how the transaction volume shifts the mid-prices through time.} 
Second, the volumes $v^i_{t^i_h}$ depend on previous volumes as a large source of the order-flow auto-correlation is found to be from order splitting \cite{TPLF2015}. Additionally, the price observations are fundamentally discrete events. Whether a continuous time representation using stochastic differential equations is sufficient remains an important open question \cite{PCEPTG2020b}.

There appears to be paucity of realistic models that are able to provide a convincingly realistic 3-tuple $(t^i_h, p^i_{t^i_h}, v^i_{t^i_h})$. Point processes like Hawkes processes provide good models for the time of the events $t^i_h$. Stochastic differential equations may provide good models for the prices $p^i_{t^i_h}$ at low-frequency, and naturally provide long-memory \cite{GJR2018}. There seems to be no consensus for a model specification for the volume of market orders as the distributions varies widely with the product and market \cite{CTPA2011a}.

Solving these issues are beyond the scope of the paper. Our objective here is to highlight various concerns and what we do empirically know about the behaviour of collections of traded asset events at high-frequency under different choices of time. For the purposes of investigating the Epps effect, the Hawkes model in \cref{eq:7} provides a good basis for the simulation because it deals with $t^i_h$ and $p^i_{t^i_h}$ in the tuple. Even though it has unrealistic price changes, the model naturally recovers microscopic properties such as the signature plot, mean reversion and the Epps effect, and it can also naturally recover large scale diffusive properties \cite{BMM2015}.

To complete the tuple, we need a method of generating $v^i_{t^i_h}$. Empirically, it was found that the unconditional distribution of market orders follows a power-law behaviour \cite{CTPA2011a}. Therefore, combining \cref{eq:7} and a power-law for the volume samples, we have a toy model to investigate the Epps effect in volume time. Even though the model does not account for the empirical form of measured price impact, we still find that it recovers the empirical dynamics surprisingly well.

In order to perform the simulation work for volume time, we will assume that the volume associated with each transaction is from an independent and identically distributed (IID) power-law where the probability density function is given as:
\begin{equation}\label{eq:14}
    f(x)=
    \begin{cases}
    \frac{\alpha x_{\mathrm{m}}^{\alpha}}{x^{\alpha+1}} & x \geq x_{\mathrm{m}} \\
    0 & x<x_{\mathrm{m}}.
    \end{cases}
\end{equation}
For the toy model, we set $x_{\mathrm{m}} = 20$ with exponent of $1 + \alpha = 2.7$ --- the value found in \citet{CTPA2011a} for market orders. Note that the volume samples are rounded to an integer.

\Cref{fig:VolumeTimePrices} demonstrates the volume time samples of the Hawkes price models with volume samples from an IID power-law distribution. \Cref{fig:VolumeTimePrices:a,fig:VolumeTimePrices:b} are 10 unit and 5 unit sampling intervals in volume time respectively. The sampling interval was chosen to be the equivalent interval in calendar time based on the number of samples. Each security has different sized volume buckets (for a given sampling interval) to achieve the synchronous and homogeneous samples. The samples here are no longer the same as those in calendar time since these samples have gone through an averaging process. This is fundamentally different to event time where the samples remain the same but smoothed.

\Cref{fig:SimVT} plots the correlation for the RV, MM and HY estimates in volume time measured at different sampling intervals. Here we see that all the estimators recover the same estimates (see \Cref{subsec:relation}).

\begin{figure}[t]
    \centering
    \includegraphics[width = 0.48\textwidth]{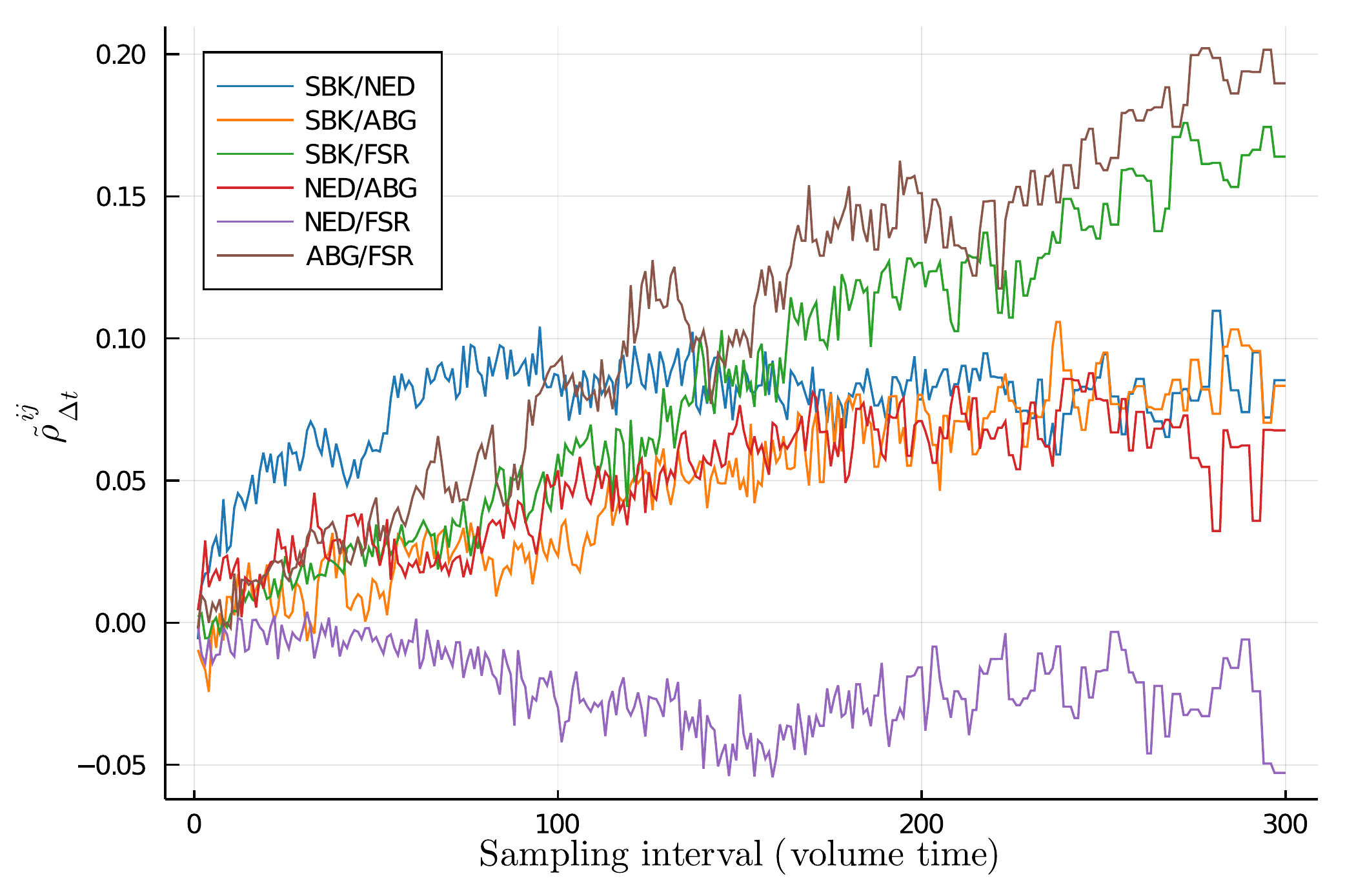}
    \caption{The Epps effect in volume time for the six correlation pairs. The sampling interval ranges from 1 to 300 units. The figure only reports the RV estimates because the three estimators recover the same estimates under volume time (see \Cref{fig:SimVT}).}
    \label{fig:EmpVT}
\end{figure}

We see that the Epps effect is present in volume time. However, something interesting happens here. The concavity of the Epps effect is lost, here the correlations emerge linearly for larger sampling intervals and there is a significant drop in the correlation achieved at larger intervals. This behaviour is not just an anomaly under simulation. We see that this is also the case with empirical data (see \Cref{fig:EmpVT}).

To try and further understand this effect, \Cref{fig:MoreVT} computes the same Epps curves in volume time as \Cref{fig:SimVT}, but for a range of distributions generating the volume samples. 

We consider the additional uniform, normal and beta distributions. The uniform distribution samples transaction volumes between 1 to 100 with equal probability. The normal distribution samples transaction volumes with a mean of 50 and standard deviation of 5. We consider three beta distributions with $\alpha = \beta$ but taking on values of 0.1, 0.2 and 2.\footnote{The beta distribution samples values between 0 and 1. Therefore, the samples are multiplied by 100 before rounding to obtain a transaction volume.} The distributions all have a mean of roughly 50 shares for each transaction volume.

We see that regardless of the distribution, the Epps effect is linear in volume time. The exact cause leading to a linear Epps curve is not clear. However, from \Cref{fig:MoreVT} we see that the linear effect does not seem to depend on the particular distribution generating the volume samples as all the distributions present this linear effect.\footnote{Unravelling potential lead-lag effects is particularly difficult for volume time because lead-lag cannot be easily separated from shrinking and stretching as these are also mixed by the averaging required to construct volume bins.}

\begin{figure}[t]
    \centering
    \includegraphics[width = 0.5\textwidth]{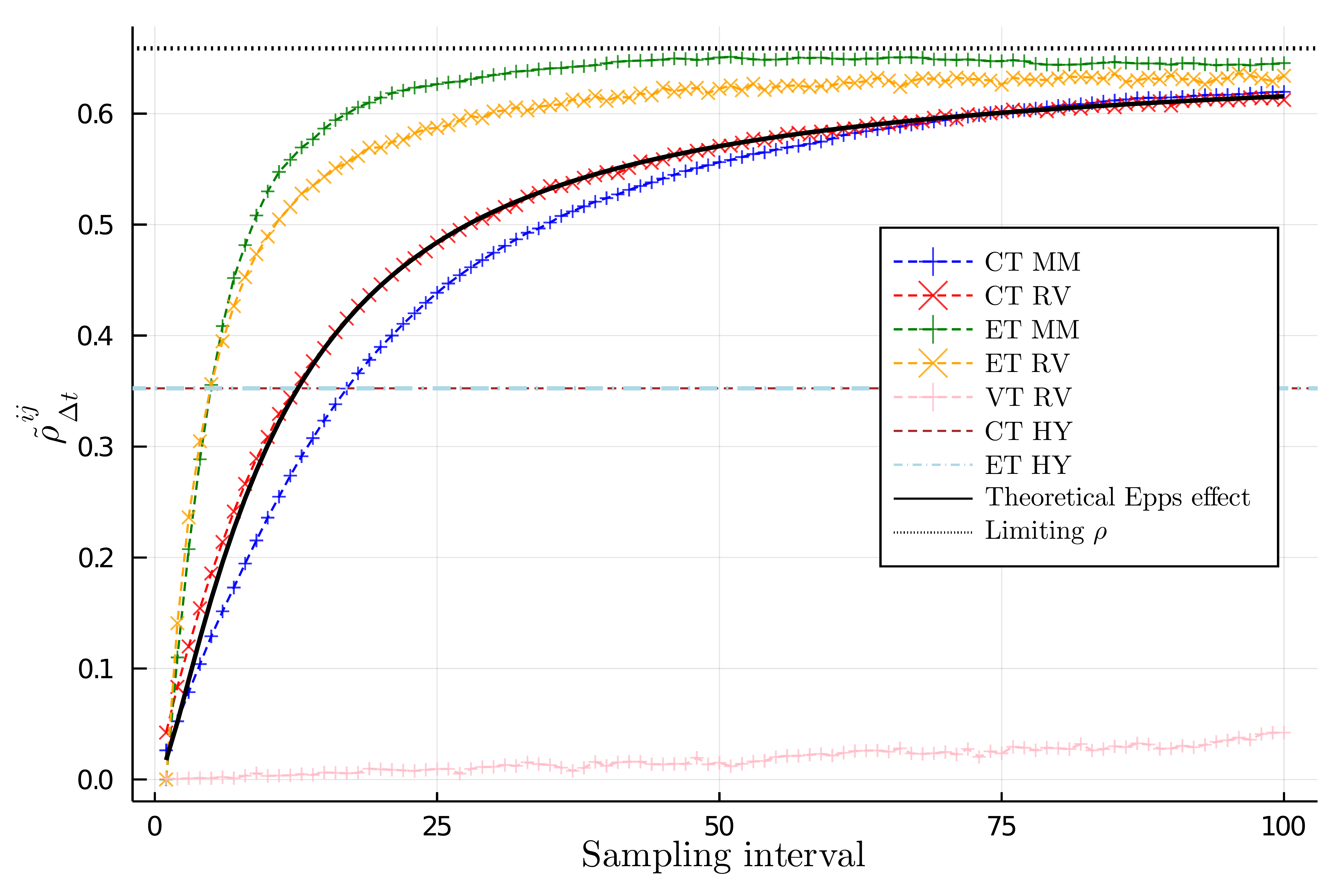}
    \caption{Comparison of the Epps curves from \Cref{fig:SimCT,fig:SimET,fig:SimVT}. We investigate calendar time (CT), event time (ET) and volume time (VT) using the RV, MM and HY estimator. The theoretical Epps effect \cref{eq:10} and limiting correlation \cref{eq:11} from calendar time are also included.}
    \label{fig:SimComp}
\end{figure}

Finally, to end this section, we compare all the Epps curves from \Cref{fig:SimCT,fig:SimET,fig:SimVT} together in \Cref{fig:SimComp}. We see that the Epps effect is present under all three definitions of time. For calendar and event time, sampling in the time domain or Fourier domain both result in an Epps effect. Interestingly, the Epps effect becomes linear under volume time.

\section{Empirical}\label{sec:emp}

We investigate the Epps effect under different definitions of time using transaction data from the Johannesburg Stock Exchange (JSE) for five trading days ranging from 24/06/2019 to 28/06/2019. We consider Standard Bank Group Ltd. (SBK), Nedbank Group Ltd. (NED), Absa Group Ltd. (ABG) and FirstRand Ltd. (FSR). The data was extracted from Bloomberg Pro and is only reported up to an accuracy of seconds (hence the data has been effectively re-sampled). Although the transactions were in the correct order, without an exact millisecond or nanosecond time stamp means that we cannot correctly order the events for event time sampling.\footnote{Consider two securities which have 3 and 5 transactions that all occurred in the same second. It is not possible to order these events on a joint event count.} Therefore, we process the data by aggregating transactions (for each security) with the same time stamp using a volume weighted average.

\begin{figure*}[htb]
    \centering
    \subfloat[MM]{\label{fig:EmpCT:a}\includegraphics[width=0.33\textwidth]{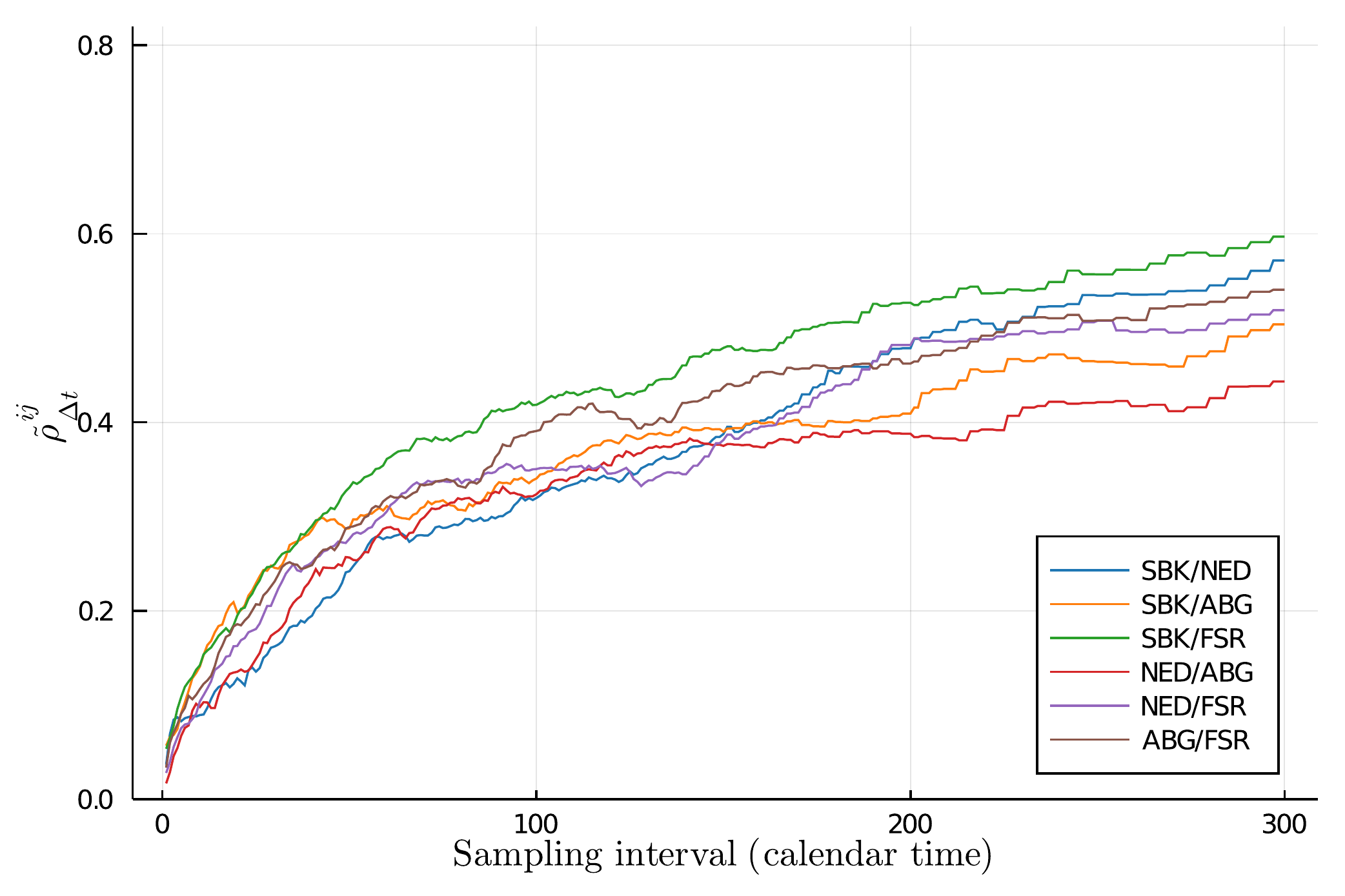}}
    \subfloat[RV]{\label{fig:EmpCT:b}\includegraphics[width=0.33\textwidth]{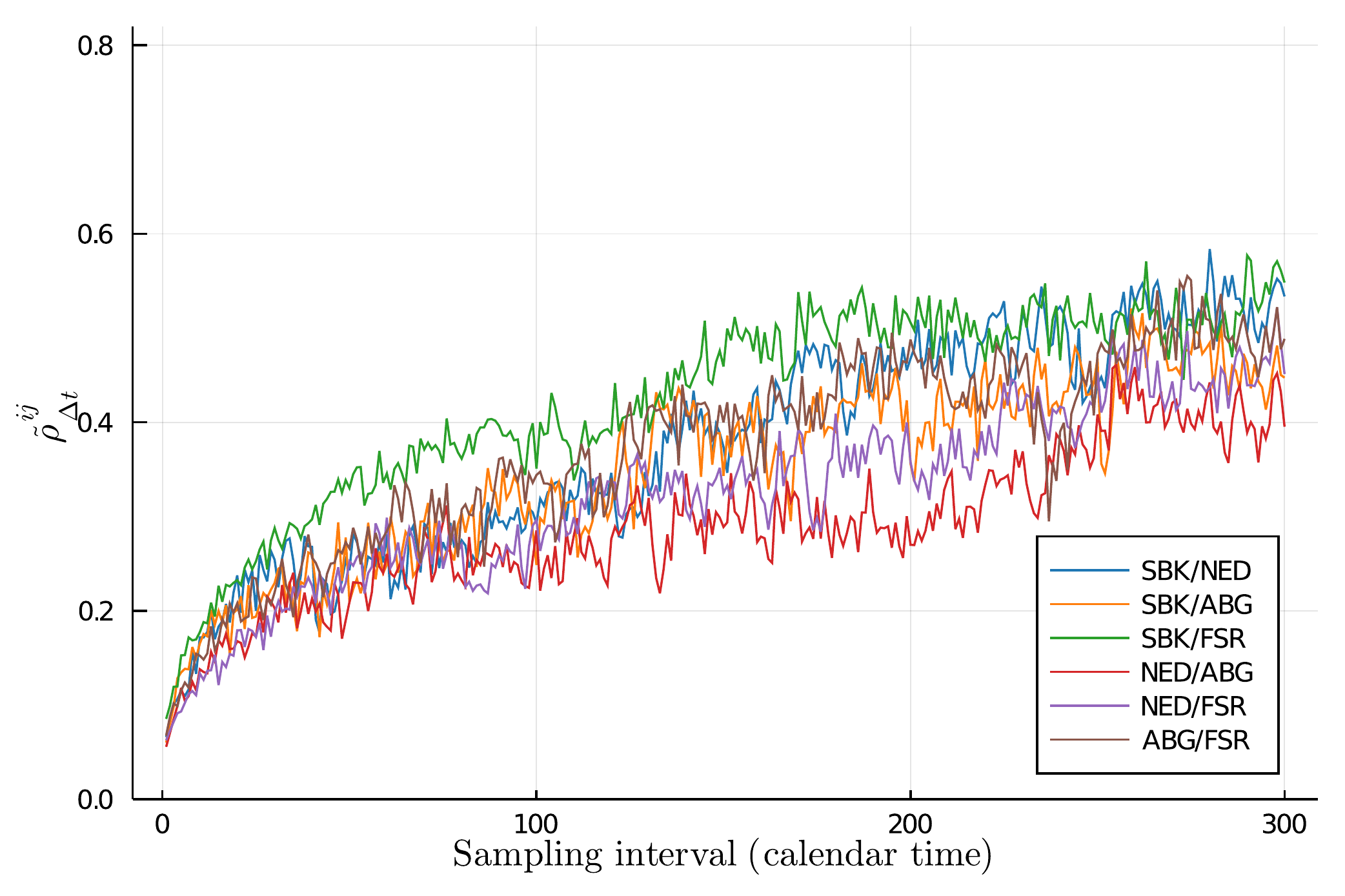}}
    \subfloat[HY]{\label{fig:EmpCT:c}\includegraphics[width=0.33\textwidth]{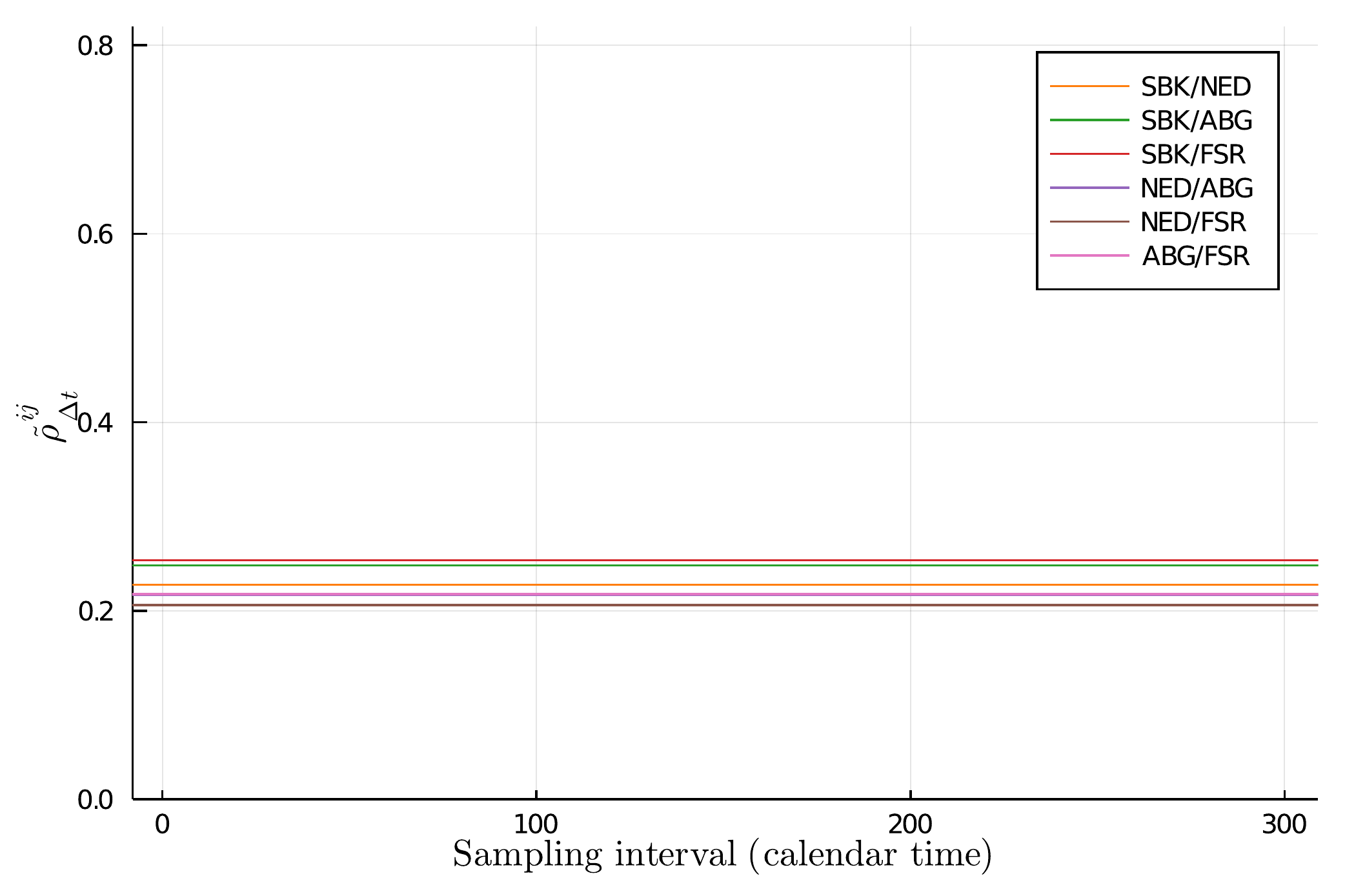}}
    \caption{The Epps effect in calendar time for the six correlation pairs. The sampling interval ranges from 1 to 300 units. \Cref{fig:EmpCT:a,fig:EmpCT:b,fig:EmpCT:c} we have the MM, RV and HY estimates respectively.}
\label{fig:EmpCT}
\end{figure*}

\begin{figure*}[htb]
    \centering
    \subfloat[MM]{\label{fig:EmpET:a}\includegraphics[width=0.33\textwidth]{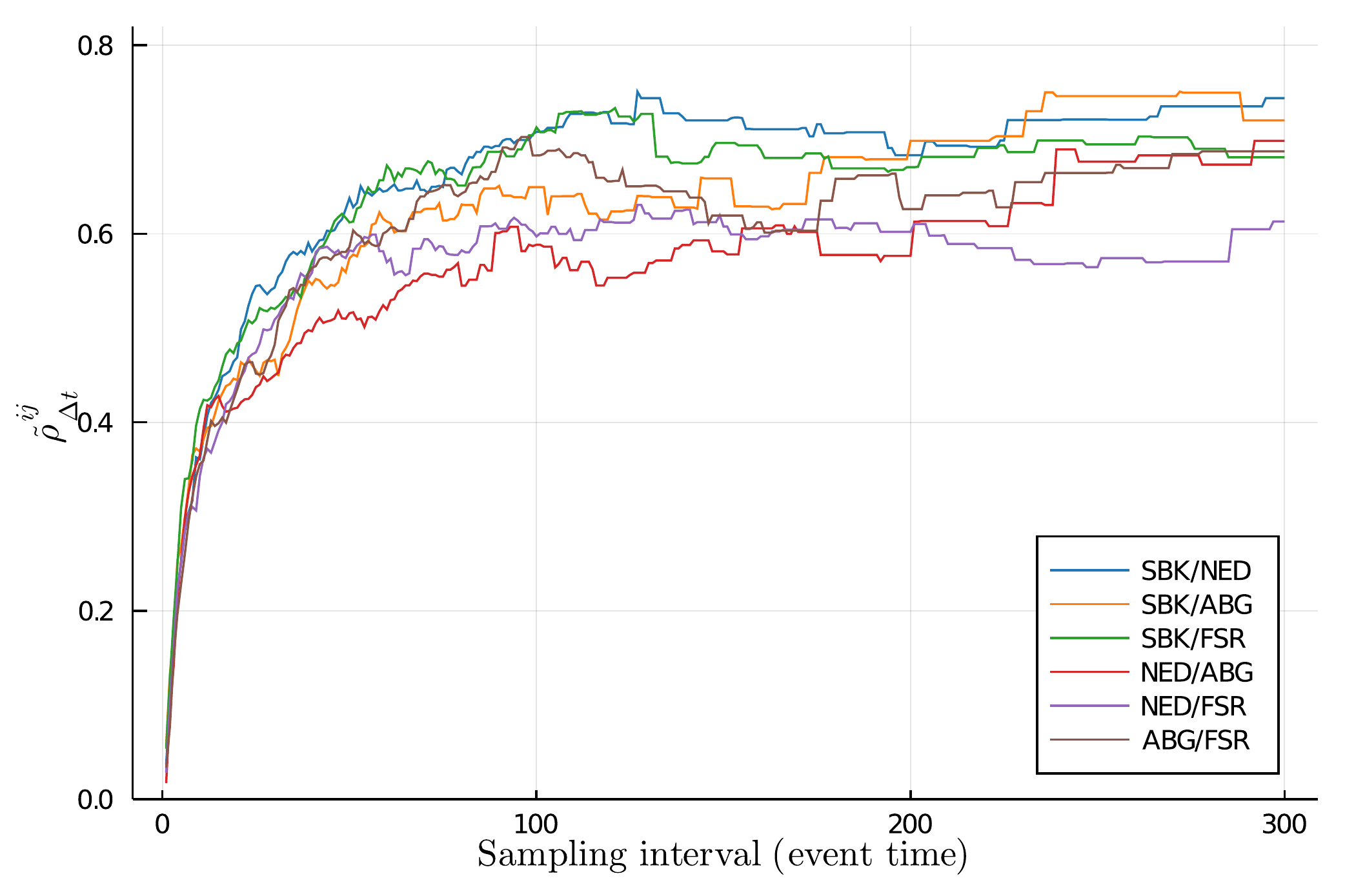}}
    \subfloat[RV]{\label{fig:EmpET:b}\includegraphics[width=0.33\textwidth]{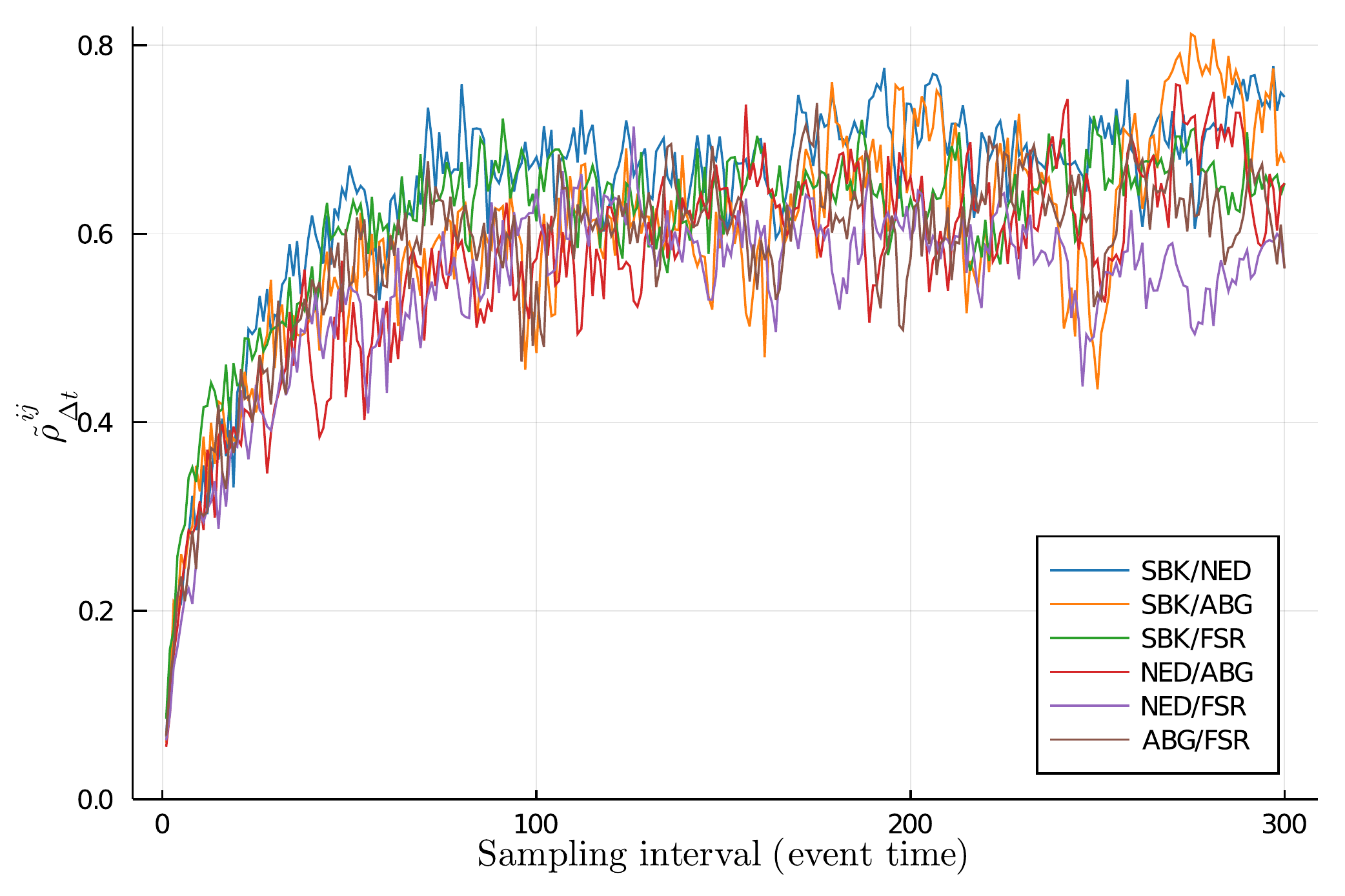}}
    \subfloat[HY]{\label{fig:EmpET:c}\includegraphics[width=0.33\textwidth]{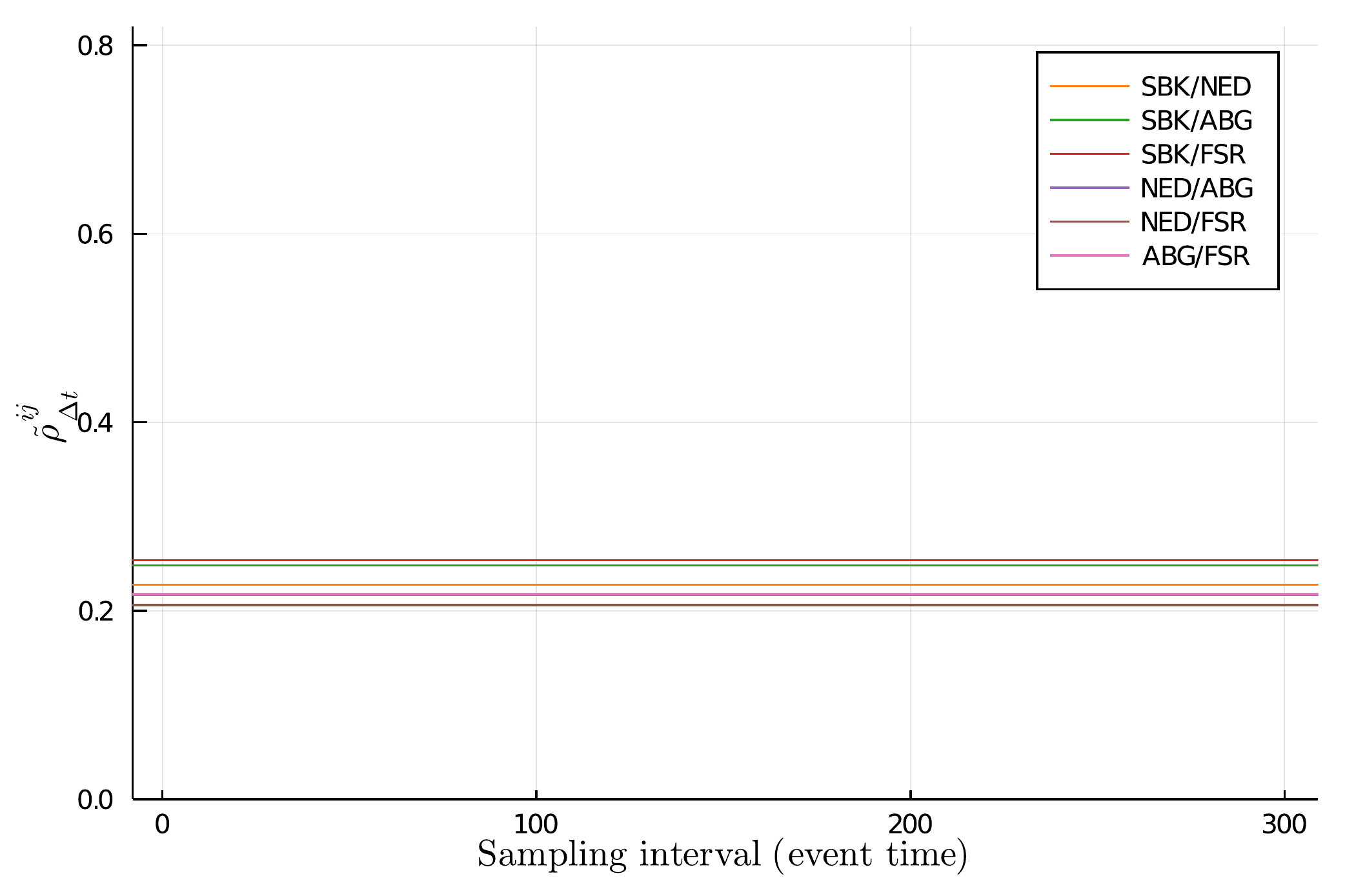}}
    \caption{The Epps effect in event time for the six correlation pairs. The sampling interval ranges from 1 to 300 units. \Cref{fig:EmpET:a,fig:EmpET:b,fig:EmpET:c} we have the MM, RV and HY estimates respectively.}
\label{fig:EmpET}
\end{figure*}

Correlations are computed separately for the trading days so that we do not have to worry about overnight returns. Therefore the correlations reported in the following subsections are the average at each sampling interval aggregated across the five trading days.

\subsection{Calendar time}\label{subsec:emp:CT}

The continuous trading session on the JSE is between 09:00 and 16:50, this translates to a seven hour and 50 minute trading day or $T=28,200$ seconds. As before, we sample a synchronous and homogeneous grid at different time scales using previous tick interpolation for the RV estimator. The MM and HY estimator use the samples from when the transactions occurred. The MM investigates different time scales by sampling the Fourier domain through the relation in \cref{eq:Der:3} while the HY provides a baseline estimate after correcting for asynchrony.

\Cref{fig:EmpCT} plots the Epps effect in calendar time for the six correlation pairs for sampling intervals ranging from 1 to 300 units for the MM, RV and HY estimator in \Cref{fig:EmpCT:a,fig:EmpCT:b,fig:EmpCT:c} respectively.

First, we see that the MM and RV estimates have yet to reach the limiting (asymptotic) correlation after 300 seconds. Second, we see that the RV estimates are less stable moving from one sampling interval to the next. This is because sampling different time scales using previous tick interpolation changes the samples (through the loss in resolution) for the estimator. This was not seen in \Cref{fig:SimCT} because this instability was hidden away when averaging 100 correlation estimates; whereas here we are only averaging over five days of trading.

\subsection{Event time}\label{subsec:emp:ET}

Lining up the events in event time with empirical data is slightly problematic. The data set we have only reports time stamps up to the second, this means that we can have concurrent events at the same time between a pair of securities (even after aggregating events on the same second using a volume weighted average). In theory, these should be two separate event counts when sharing the same event clock, however since there is no way to determine the ordering of these events, we allow them to happen simultaneously, {\it i.e.,} they share the same event count. This was not a problem under simulation because we had a well-defined point process.

\Cref{fig:EmpET} plots the Epps effect in event time for the six correlation pairs for sampling intervals ranging from 1 to 300 units for the MM, RV and HY estimator in \Cref{fig:EmpET:a,fig:EmpET:b,fig:EmpET:c} respectively.

We see again that the Epps effect is present and correlations seem to emerge faster in event time. Both the MM and RV estimator reaches the limiting (asymptotic) correlation after around 100 units in event time. Again, we also see that the HY estimate is exactly the same as the estimate in calendar time. This is due to how the events are lined up. Even though the time series is smoothed out in event time, the observations in \cref{eq:Der:4} remain the same and the intervals where $w_{h \ell}$ is activated remains the same.

\begin{figure*}[htb]
    \centering
    \subfloat[SBK/FSR]{\label{fig:EmpComp:a}\includegraphics[width=0.5\textwidth]{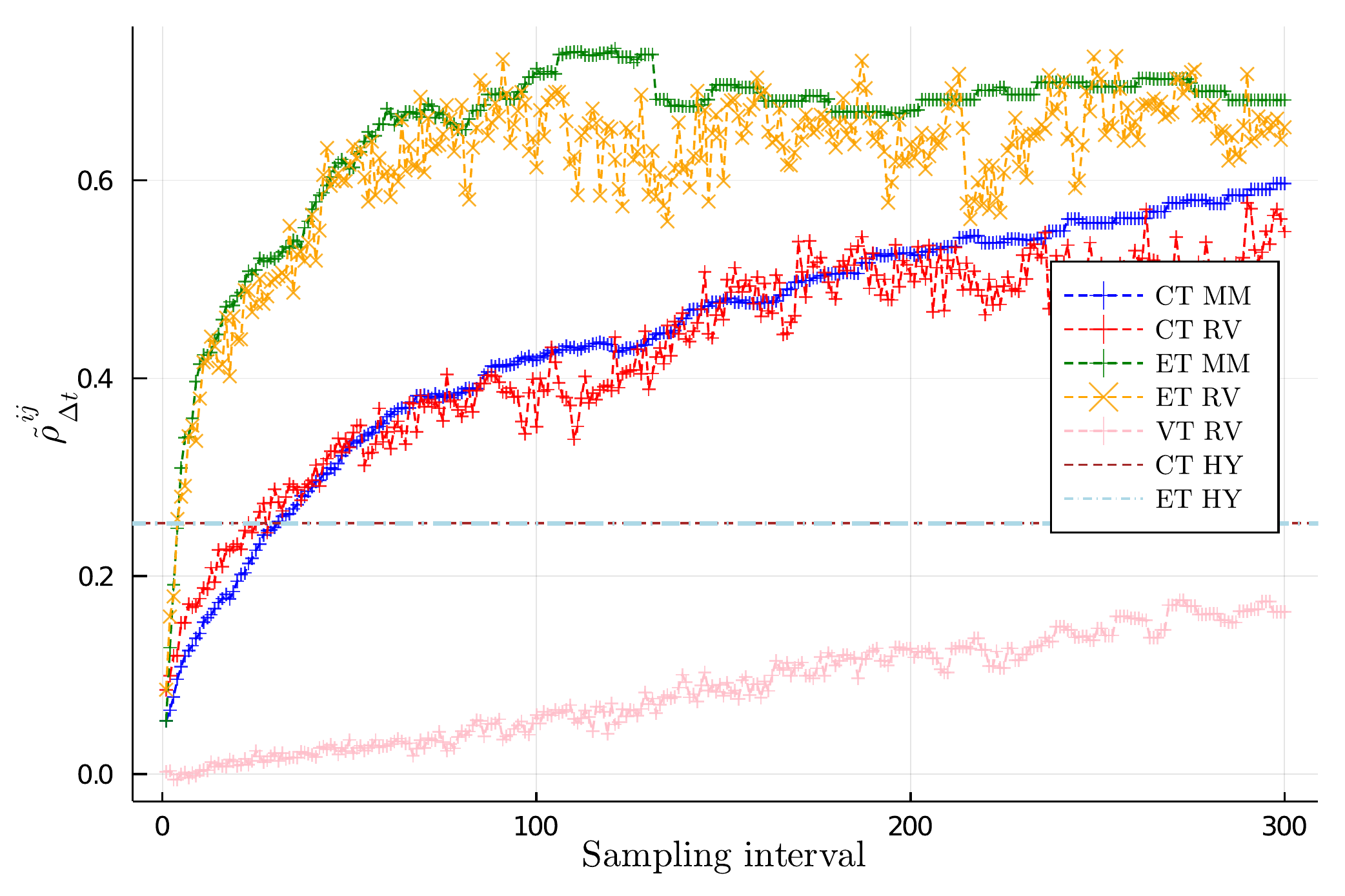}}
    \subfloat[NED/ABG]{\label{fig:EmpComp:b}\includegraphics[width=0.5\textwidth]{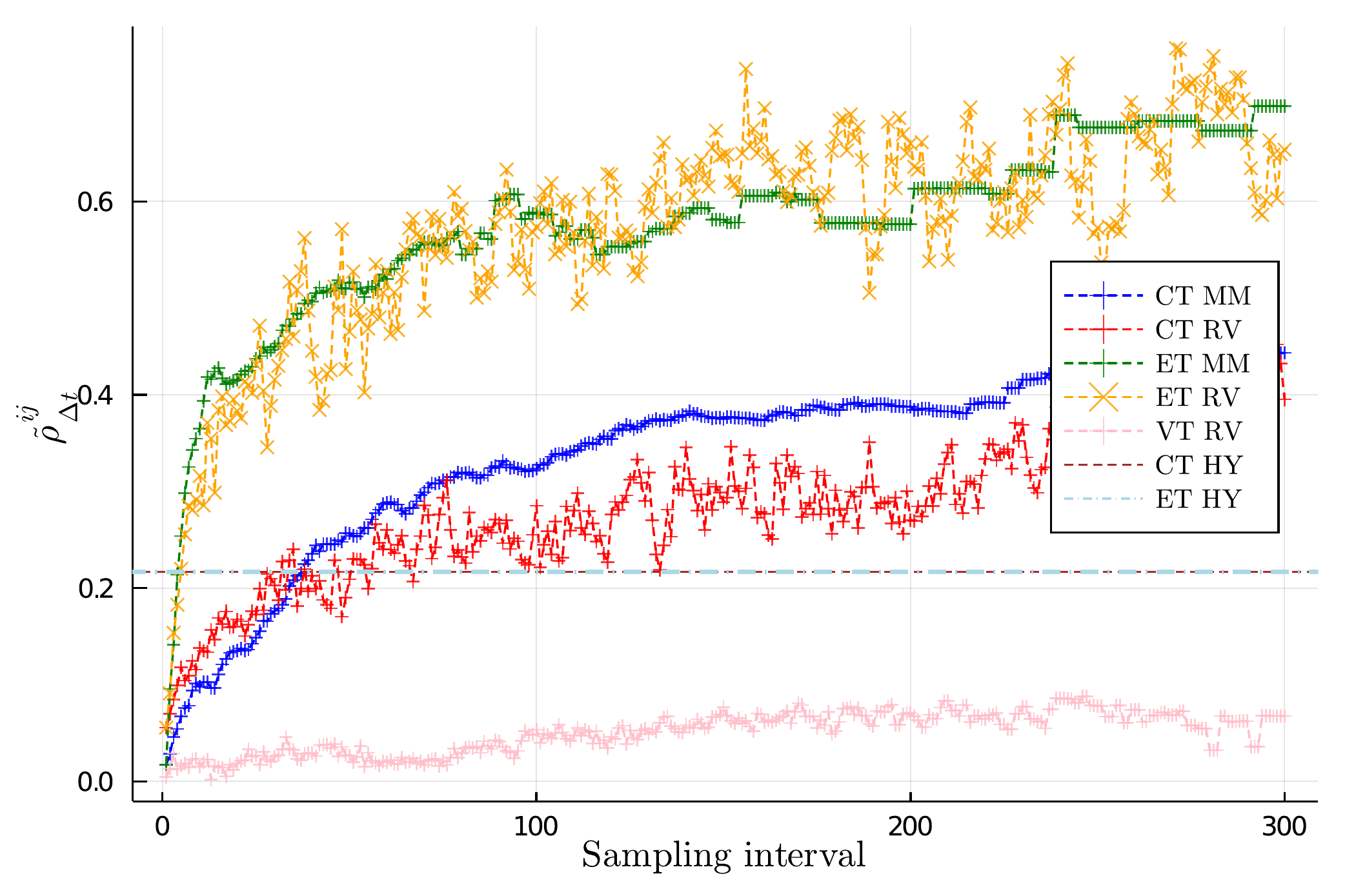}}
    \caption{The Epps curves compared under different definitions of time for (a) the SBK/FSR and (b) the NED/ABG pair. Included are the calendar time (CT), event time (ET) and volume time (VT) estimates for the range of 1 to 300 unit sampling intervals using the RV and MM estimator. The HY estimator is provided as a baseline.}
\label{fig:EmpComp}
\end{figure*}

\subsection{Volume time}\label{subsec:emp:VT}

Sampling in volume time is done by aggregating transaction prices over a certain number of shares determined by the volume bucket. As before, the sampling interval here is defined by the number of samples obtained. The number of samples obtained is compared against the number of sampled obtained in calendar time to find an equivalent sampling interval in volume time.

\Cref{fig:EmpVT} plots the Epps effect in volume time for the six correlation pairs for sampling intervals ranging from 1 to 300 units. Here we only report the RV estimates because the three estimators recover the same estimates in volume time (see \Cref{fig:SimVT}).

We see that as in the case of our simulations, the Epps effect is present in volume time. Moreover, correlations emerge linearly rather than exponentially for larger sampling intervals as in the case of calendar and event time; and there is a significant drop in the correlations achieved for larger sampling intervals compared calendar and event time.

Looking at the NED/FSR pair (purple line) in \Cref{fig:EmpVT}, we see that the correlation becomes negative in volume time. This behaviour is possible under simulation with our toy model as the error bars in \Cref{fig:SimVT} do go into the negative region, meaning that this is possible. However, at this stage it remains unclear as to why this is occurs.

\subsection{Comparison}\label{subsec:emp:comp}

\Cref{fig:EmpComp} compares the various the Epps curves under different definitions of time. This is done for two of the six correlations pairs SBK/FSR and NED/ABG in \Cref{fig:EmpComp:a,fig:EmpComp:b} respectively.

Comparing \Cref{fig:EmpComp} against \Cref{fig:SimComp}, we see that the empirical behaviours of the various Epps curves line up with what we have seen using the toy model under simulation. The event time correlations emerge faster than those in calendar time, while the Epps effect is linear under volume time. Sampling in the time domain or Fourier domain both result in an Epps effect under calendar and event time. In \ref{app:A}, we provide the same comparison between uncorrelated asset pairs. The generalisation of these results is only confirmed to hold for correlated asset pairs where the Epps effect behaves as expected.

Given that this relatively simple toy model captures the various empirical dynamics, it is unlikely that a more realistic simulation of the 3-tuple $(t^i_h, p^i_{t^i_h}, v^i_{t^i_h})$ would affect the correlation estimates. In particular, intraday seasonality of the volatility, transactions, and volumes that are present with empirical data \cite{PCJ2015} does not alter the dynamics compared to the toy model which does not capture these seasonality effects. A possible reason why the intraday seasonality effects do not alter the dynamics is because the correlation estimates are integrated quantities.\footnote{We speculate that intraday seasonality may alter the dynamic of the correlations in volume time, particularly when two assets have significantly different intraday volume curves.} Perhaps these seasonality effects may influence the dynamics differently if instantaneous correlation quantities were computed instead.

\section{Conclusion}\label{sec:conc}

In this paper, we compared the Epps effect under three definitions of time: calendar time, trade time and volume time. We use a Hawkes process, specifically, the fine-to-coarse model by \citet{BDHM2013a} as the basis for the simulation of price paths. Combining this with a power-law distribution for the volumes samples \cite{CTPA2011a}, we find that the Epps effect is present under all three definitions of time. 

Moreover, the simulations reveal different rates of correlation emergence. First, we saw that the correlations emerge faster under trade time compared to calendar time. Second, correlations emerge linearly under volume time and do not seem to depend strongly on the distribution generating the volume samples. Under this setup, where the volumes and prices are independent, we find realistic results. These results are found despite using a model that does not directly encode price impact or intraday seasonality of the volatility, transactions, and volumes.

We then investigated the Epps effect under these three definitions of time using transaction data from the Johannesburg Stock Exchange. We found that the empirical Epps curves conform well with what we observed under simulation for the various definitions of time. However, the results hold for the case of sufficiently correlated asset pairs where the Epps effect behaves as expected.

This paper presents some anomalies that raise several unresolved concerns. First, the underlying mechanisms that link the different temporal definitions and sampling schemes that lead to the different correlation dynamics remains unclear. Second, an analysis of the instantaneous correlation estimates could be performed for the different definitions of time. This could potentially lead to a better understanding of how intraday seasonality could impact the correlation dynamics under different sampling schemes. Finally, to explore how market activity affects the Epps effect with different time choices. This particularly relates to the observation made by \citet{TK2007} where the Epps curves do not scale with market activity.

\section*{Reproducibility of the Research}
The dataset can be found at \cite{PCEPTG2020DATAc}. The Julia code and instructions for replication can be found in our GitHub site \cite{PCEPTG2020CODEc}.

\section*{Acknowledgements}
We would like to thank the two anonymous referees for thoughtful comments which helped us improve this paper.
PC would like to thank Roger Bukuru for work done together and various discussions.

\section*{Funding}
PC was supported by the Manuel \& Luby Washkansky Scholarship and the South African Statistical Association [grant number 127931]. 

\balance
\bibliographystyle{elsarticle-harv}
\bibliography{PCEPTG-AltSampling.bib}

\onecolumn
\appendix

\section{Uncorrelated assets}\label{app:A}

\begin{figure*}[htb]
    \centering
    \subfloat[NED/AGL]{\label{fig:EmpAppComp:a}\includegraphics[width=0.5\textwidth]{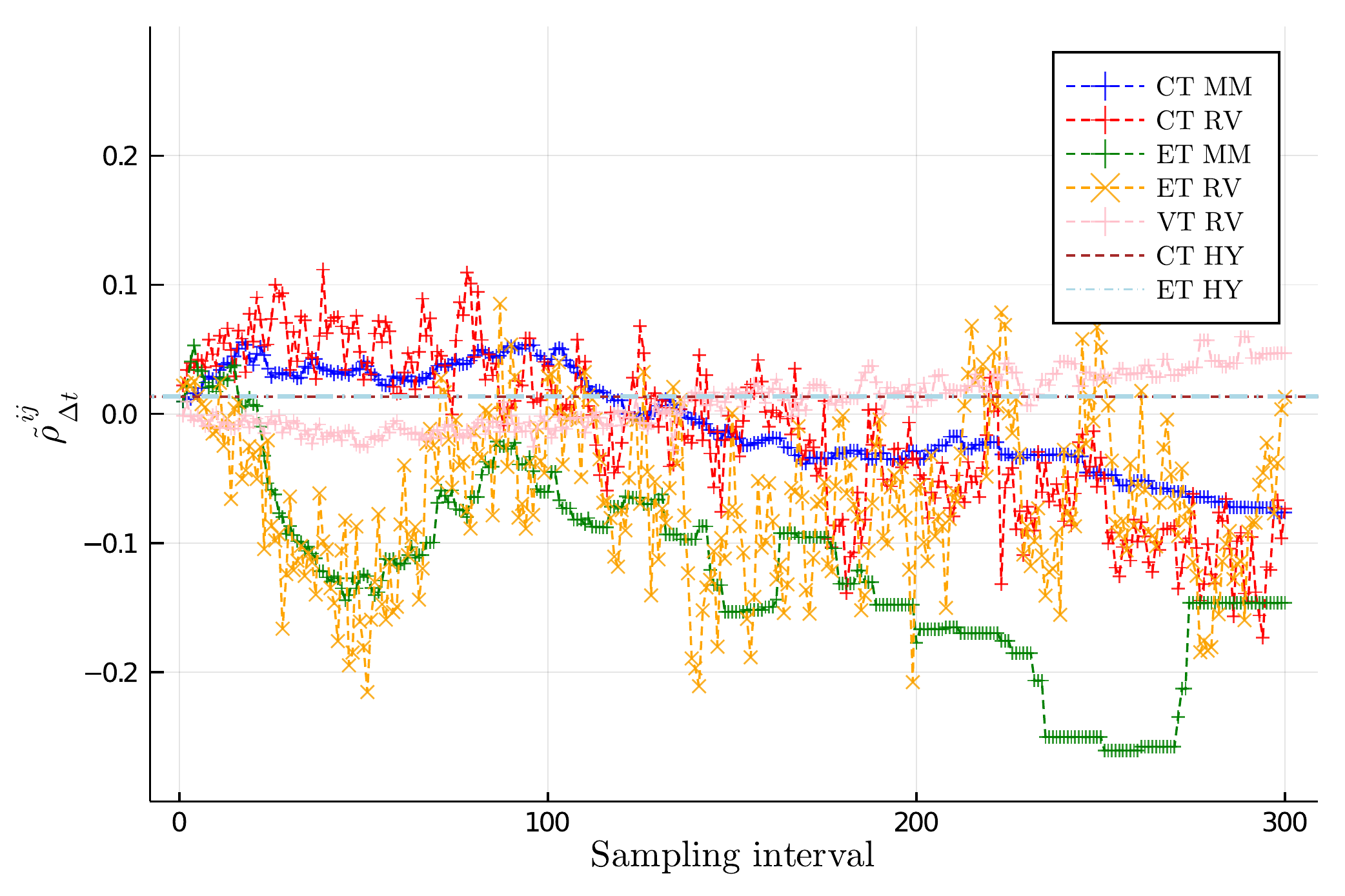}}
    \subfloat[NED/BTI]{\label{fig:EmpAppComp:b}\includegraphics[width=0.5\textwidth]{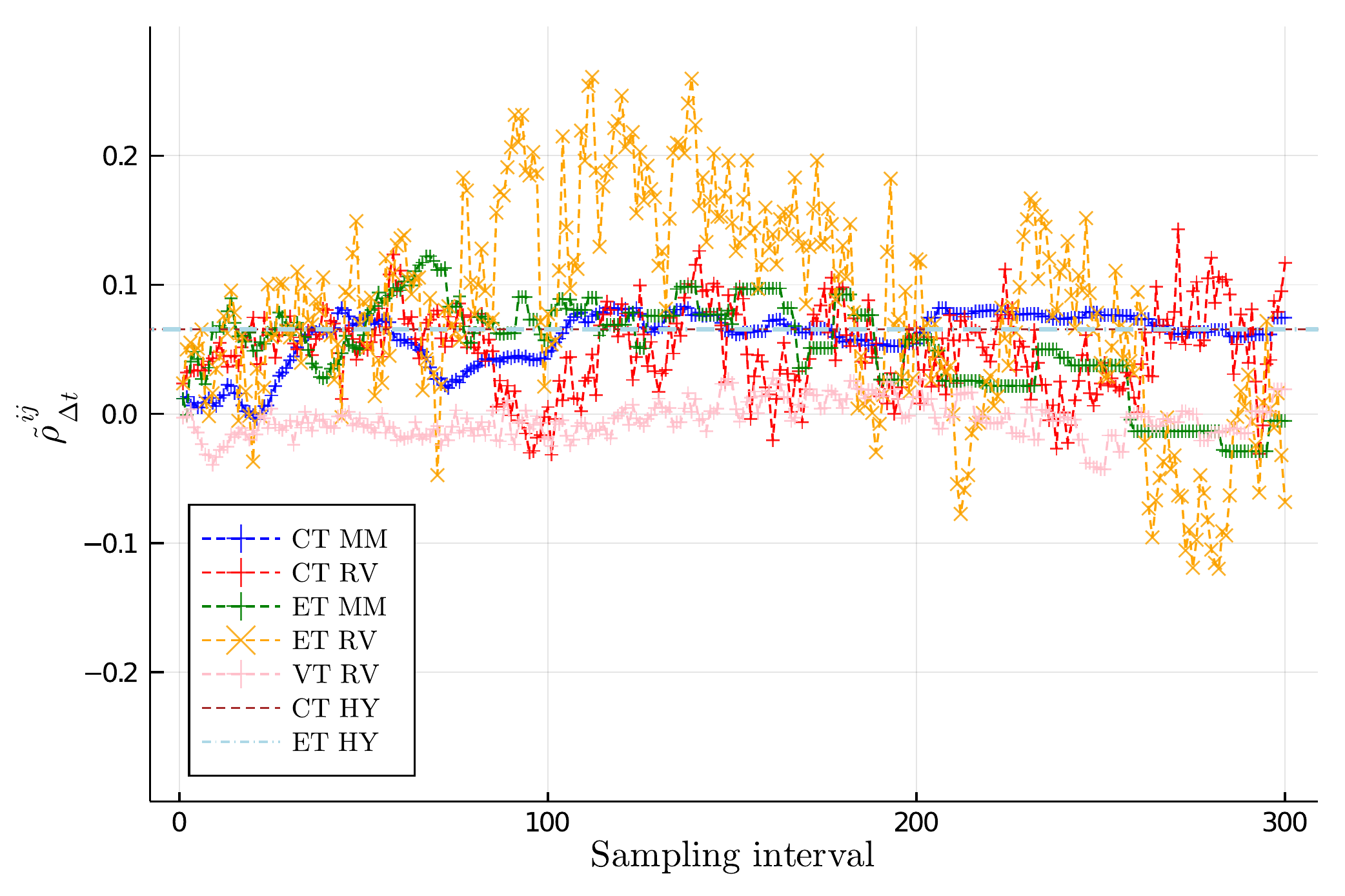}}
    \caption{The Epps curves compared under different definitions of time for (a) the NED/AGL and (b) the NED/BTI pair. Included are the calendar time (CT), event time (ET) and volume time (VT) estimates for the range of 1 to 300 unit sampling intervals using the RV and MM estimator. The HY estimator is provided as a baseline.}
\label{fig:EmpAppComp}
\end{figure*}

We investigate whether the correlation dynamics found between the different definitions of time hold in general for all asset pairs. To this end, \Cref{fig:EmpAppComp} compares the Epps curves under different sampling schemes for the asset pairs (a) NED and Anglo American plc (AGL), and (b) NED and British American Tobacco plc (BTI). The correlations are reported in the same way as \Cref{sec:emp} by computing the correlations separately for the five trading days and reporting the average at each sampling interval.

From the figure it is suggested that the Epps effect is not prominent in the case of uncorrelated assets --- this is an unsurprising result. However, we still see that correlations emerge linearly and that there is are drop in the correlations achieved in volume time compared to calendar and event time. The correlation dynamics between calendar and event time becomes ambiguous. 

Here the Epps effect itself starts to behave in an unexpected manner. The calendar time correlations for the NED/AGL pair start off positive and become negative for larger sampling intervals. A similar dynamic of the Epps effect in calendar time has been reported between the FSR/AGL pair in \citet{PCEPTG2020a}. However, to the best of our knowledge, this has not been reported in the boarder literature surrounding the Epps effect. This is likely because the literature generally focuses on asset pairs with higher correlation levels where the Epps effect then behaves as expected. Therefore, we conclude that our results only hold in the case of sufficiently correlated assets. We speculate that this is due to the nature of different types of trading agent activity on different time scales.

\end{document}